\documentclass[a4paper,fleqn,usenatbib]{mnras}

\usepackage{newtxtext,newtxmath}
\usepackage[T1]{fontenc}
\usepackage{ae,aecompl}

\usepackage{graphicx}	% Including figure files
\usepackage{amsmath}	% Advanced maths commands
\usepackage{amssymb}	% Extra maths symbols

\title{Triaxial instabilities in rapidly rotating Neutron Stars.}

\author[A. Basak]{Arkadip Basak$^{1}$\thanks{E-mail: \href{mailto:arkadip.basak@cfa.harvard.edu}{arkadip.basak@cfa.harvard.edu}}
\\
$^{1}$Harvard Smithsonian Center for Astrophysics,
60 Garden Street, Cambridge, MA 02138.}

\date{Accepted XXX. Received YYY; in original form ZZZ}

\pubyear{2017}

\begin{document}
\label{firstpage}
\pagerange{\pageref{firstpage}--\pageref{lastpage}}
\maketitle

\begin{abstract}
Viscosity driven bar mode secular instabilities of rapidly rotating neutron stars are studied using LORENE/Nrotstar code. These instabilities set a more rigorous limit to the rotation frequency of a neutron star than the Kepler frequency/mass-shedding limit.  The procedure employed in the code comprises of perturbing an axisymmetric and stationary configuration of a neutron star and studying its evolution by constructing a series of triaxial quasi-equilibrium configurations. Symmetry breaking point was found out for Polytropic as well as 10 realistic Equations of states (EOS) from the CompOSE database. The concept of piecewise polytropic EOSs has been used to comprehend the rotational instability of Realistic EOSs and validated with 19 different Realistic EOSs from CompOSE. The possibility of detecting quasi-periodic gravitational waves from viscosity driven instability with ground-based LIGO/VIRGO interferometers is also discussed very briefly.
\end{abstract}


%%%%%%%%%%%%%%%%%%%%%%%%%%%%%%%%%%%%%%%%%%%%%%%%%%

%%%%%%%%%%%%%%%%% BODY OF PAPER %%%%%%%%%%%%%%%%%%

\section{Introduction}
\label{sec:intro}
Newborn rotating neutron stars can spontaneously break their axial symmetry if the ratio
of the kinetic energy $T$ to that of the absolute value of
the gravitational potential energy $W$, exceeds some threshold value
\citep[]{eric1, eric2}. Similarly, an evolved neutron star in a closed binary
system, while accreting matter from its companion may acquire a high enough angular momentum reach a stage where the
ratio $T/W$ is large enough to allow for this symmetry breaking. Moreover, \cite{cite_new1} also suggests that supernova fallback accretion can spin up a newly formed strongly magnetized neutron star to high values of $T/W$. A steady-state configuration is achieved at an instant when the total accreted
angular momentum is expelled via gravitational radiation \cite[]{eric1},
by a process known as forced gravitational emission. However, the energy radiated by a newborn neutron star is fuelled by its rotational kinetic energy, while for an accreting star, it is provided by the accreted matter\cite[]{cite46}. Also, a newborn neutron star is way more hot than the accreting one. Thus, the viscosity differs considerably between the two cases, which in turn causes the instability generating the symmetry breaking to be different. There are two kinds of instabilities which can set into rapidly rotating stars \cite[]{cite40}, one driven by gravitational radiation reaction, also known as the Chandrasekhar-Friedman-Schutz (CFS) instability \cite[]{cite39, cite39_3, cite41, cite42}, while the other driven by viscosity \cite[]{cite43}.

%. The accretion rate of the fallback material into the neutron star can be as high as $\dot{M} \simeq 10^{-4} - 10^{-2} M_{\odot}$ thereby transporting enough angular momentum to enable symmetry breaking.

A self gravitating incompressible fluid rotating rigidly at some moderate velocity, takes up the form of an axisymmetric ellipsoid, known as the Maclaurin spheroid \cite[]{cite41}. At critical value of $T/W = 0.1375$ \cite[]{eric1}, the two families of triaxial ellipsoids separate: the Jacobi ellipsoids and the Dedekind ellipsoids. Maclurin spheroids are dynamically unstable for $T/W \geq 0.2738$ \cite[]{eric2}. This implies that the Jacobi/Dedekind bifurcation point $T/|W| = 0.1375$ \cite[]{eric2} is stable while rotating. However, in the presence of some dissipative forces like viscosity or gravitational radiation reaction (CFS instability), the bifurcation point tend to be always unstable at $ l = 2, m = 2$ bar mode \cite[]{eric1, citeJ} \footnote{Refer to \cite{cite3, citeJ} for more details on bar modes.}. 

A non-dissipative mechanism such as a magnetic field with a component parallel to the rotation axis breaks circular conservation \cite[]{cite44} and introduces spontaneous symmetry breaking. If only the viscosity is considered, the growth of the bar mode leads to the deformation of Maclaurin spheroids following the Riemann S ellipsoids \cite[]{cite39_1, eric1, cite46} sequence ultimately concluding as a Jacobi ellipsoid. On the contrary, if only the gravitational radiation reaction is considered instead of viscosity, the Maclaurin spheroids evolves following another Riemann S Sequence ending as a Dedekind ellipsoid\cite[]{cite44}. For the first sequence, a Jacobi ellipsoid has lesser mechanical energy $T + W$ than a Maclaurin ellipsoid but with the same angular momentum and rest mass. Since viscosity dissipates mechanical energy but conserves angular momentum, the evolution from a Maclaurin to a Jacobi ellipsoid occurs with the same angular momentum and mass at $T/W \geq 0.1375$\cite[]{eric1, cite46}. At the final stage, the Jacobi ellipsoid is rigidly rotating. Hence, dissipation of mechanical energy due to viscosity recedes. The transition towards the Jacobi ellipsoid on the viscosity driven instability time scale is much longer when compared to the CFS instability \cite[]{eric2, cite46}. 

For the second case i.e. CFS instability, emission of gravitational radiation is not accompanied by the conservation of angular momentum but conservation of the fluid circulation around the star when traversed along a closed contour along a plane parallel to the equator\cite[]{eric1, cite46}. For a given circulation and (rest) mass, the value of mechanical energy for a Dedekind ellipsoid is lower than that of a Maclaurin ellipsoid. The evolution is stalled at a Dedekind ellipsoid\cite[]{eric2}, because such a body doesn't emit any gravitational radiation. This is because such a body is stationary in the inertial frame at rest with respect to the center of the star and its mass quadrupole moment doesn't vary\cite[]{eric1, cite46, cite35}. If both viscosity and gravitational radiation reaction are considered, their cumulative effect tends to stabilize the star \cite[]{cite45} as the individual effects cancel each other\cite[]{eric1, cite46}. When the two dissipative forces are exactly equal, the Maclaurin ellipsoid is stable, up to the dynamical instability point i.e. $T/W = 0.2738$. If one of the dissipative mechanisms is negligible with respect to the other, the critical value of $T/|W|$ is slightly higher than 0.1375 \cite[]{cite46}. 

These results can be applied to compressible fluids modeled by a polytropic equation of state. Bifurcation point for triaxial configurations only if adiabatic index $\gamma > \gamma_{crit} \simeq 2.2$ \cite[]{cite47, cite48}. This can be attributed to the fact that EOSs must be stiff enough for bifurcation to occur at an angular velocity lower than that of maximum angular velocity ($\Omega_{k}$) \cite[]{cite8, cite14} for which a stationary solution exists\cite[]{cite47}. $\Omega_{k}$ is reached when the centrifugal force is exactly equal in magnitude to the gravitational potential energy at the equator of the star. $\Omega_{k}$ is also known as the Keplerian velocity (mass-shedding limit) \cite[]{cite4}.

Realistic EOSs are quite different from being polytropic, primarily because they are way softer in the outer layers than in the core of the star \cite[]{eric1}. This can have some considerable impact on the value of the Keplerian velocity $\Omega_{k}$ and hence the existence of a bifurcation point. Thus, the most important question that is to to be asked is that are realistic EOSs of nuclear matter stiff enough for the bar mode $l = 2, m = 2$ \cite[]{cite3, citeJ} instability to exist for rotating neutron stars. 

Neutron stars are relativistic objects and Newtonian approximation \cite[]{eric_lecture} is not sufficient to describe Neutron stars with masses greater than $1 M_{\odot}$. A study of instabilities for Newtonian Polytropes is undertaken in \cite{eric1} and \cite{eric2} while such instabilities have been studied in full general relativity in \cite{citeJ} and \cite{cite55}. This article briefly describes the results obtained for Newtonian and Relativistic polytropes before divulging into the analysis of instabilities for realistic EOSs. 

For the analysis of Maclaurin to Jacobi bifurcation point in fully general relativistic frame, the rotation of these objects is assumed to be rigid, an idea which has been well defined for the relativistic scenario. Thus, the results in the subsequent sections are applicable if there is some efficient mechanism to rigidify the motion. This is primarily accomplished by viscosity but can also happen in the presence of a magnetic field \cite[]{eric1, cite_new2}.

This article is organized in the following manner. In
Sec.~(\ref{sec:Theoretical Model}), a brief overview of the theoretical model is
presented, including the perturbative treatment of triaxial
deformations. In Sec.~(\ref{sec:Lorene}) a brief summary of the
numerical method is provided, followed by some code
tests using analytical polytropic EoS. Results with realistic EoS will
be discussed in Sec.~(\ref{sec:EOS}). In Sec.~(\ref{sec:comp}), the
stability of realistic EOSs with respect to triaxial deformations are analyzed.

\section{Theoretical Model}
\label{sec:Theoretical Model} 

Neutron stars rotate rigidly if the viscosity is high enough to
damp out deviations from the uniform rotation. If this happens, the CFS
instability (Dedekind like mode) is prevented and the instability
follows along the Jacobi-like mode\cite[]{eric1, cite46}. The ratio of the gravitational radiation reaction to that of the strength of the viscous force governs the race between these two modes\cite[]{eric1}.

For, accreting systems at lower temperatures $(T \sim 10^6 K)$ the Dedekind-like mode (CFS instability) is
blocked by shear viscosity arising out of electron-electron
scattering \cite[]{cite49}, whereas for higher temperatures $(T \simeq
10^9 K)$, it is prevented by bulk viscosity resulting out of direct
URCA processes \cite[]{cite49, cite50}, provided that the proton
fraction is above $\sim 10\%$. For a newborn star, at temperatures
around $T \sim 10^{10} K $, neither shear nor bulk viscosity is
sufficient to prevent the Dedekind-like mode. However, it can be
prevented by the presence of a magnetic field
\cite[]{cite37}. If infinite conductivity is assumed, there is a
toroidal magnetic field generated by the shear of the fluid from
differential angular velocity, in addition to the one at the
poles. Thus, the excess kinetic energy contained in the differential
rotation with respect to the rigid rotation is efficiently converted
into magnetic energy thereby enforcing rigid rotation \cite[]{cite37}. Although a recent work \cite[]{cite_new2}, proposes a more stringent limit on the effects of magnetic field on damping out the r-mode (CFS) instability, a very high value of the saturation amplitude\footnote{Refer to section 2b of \cite{cite_new2} for the mathematical definition of $\alpha$.} ($\alpha$) can still contribute to magnetic field generation in nascent neutron stars.

The theoretical model used for the numerical code has been described at length in \cite{eric_lecture}, \cite{eric1}, \cite{eric2}, \cite{cite55}, \citep{cite36} and \cite{citeJ}. However, a brief overview of the model used in the code has been provided.

Initially, the components of the metric tensor for a rotating star using spherical coordinates $(t,r,\theta, \phi)$ are written,
\begin{multline}
    g_{\alpha \beta} dx^\alpha dx^\beta = -N^2dt^2 + B^2r^2 \sin^2 \theta (d\phi - N^\phi dt)^2 + \\ A^2(dr^2 + r^2d\theta^2).
    \label{eqn:eq1}
\end{multline}
The four functions, $N, N^\phi, A$ and $B$ depend on the coordinates $(r, \theta)$. Once the axisymmetry of the star \cite[]{cite38, cite36_1} is broken, the stationary nature of the spacetime ceases to exist. According to Newtonian theory, an inertial frame in which the rotating triaxial body appears stationary i.e. doesn't depend on time doesn't exist. It can only be stationary in a co-rotating frame, which isn't an inertial one. However, in a general relativistic case, a rotating triaxial system is not stationary, primarily because it radiates energy in the form of gravitational waves\cite[]{eric1, cite46}. Even if a co-rotating frame has been defined, the body would not be in a steady state in such a frame as it gives out energy and hence loses angular momentum through gravitational radiation \cite[]{eric2}. 

However, when the point at which symmetry breaking occurs in being considered, there has been no emission of gravitational radiation up to that point\cite[]{cite46, eric2}. For small deviations from the axisymmetric configuration, if gravitational radiation is neglected and rigid rotation is considered, there exists a killing vector field $l$ which is proportional to the 4-velocity of the fluid.
\begin{equation}
    u = \lambda l.
    \label{eqn:eq2}
\end{equation}
Here, $\lambda$ is a positive scalar function. The spacetime exhibits some helical symmetry due to the killing vector $l$ \cite[]{eric1, cite36}. In the numerical code, the metric given in equation (\ref{eqn:eq1}) is retained with the addition of no extra diagonal term apart from $g_{t\phi}$ because several gravitational potential terms have contributions from matter sources that usually depend on fluid pressure terms or the product of matter density and velocity. However, the metric elements are not functions of $(r,\theta)$ but $(r ,\theta, \psi)$, where 
\begin{equation}
    \psi = \phi - \Omega t.
    \label{eqn:eq3}
\end{equation}
$\Omega$ is the angular velocity of the star given by $\Omega = u^\phi/u^t$. In the revised co-ordinate system $(t, r, \theta, \psi), t$ is the co-ordinate associated with the Killing vector $l$, where $l = \partial/\partial t$, and $(r, \theta, \psi)$ are held fixed. The metric coefficients can now be explicitly written as:
\begin{multline}
        g_{\alpha \beta} dx^\alpha dx^\beta = -N (r, \theta, \psi)^2dt^2 + \\ \frac{\tilde{B} (r,\theta)^2}{N (r, \theta, \psi)}r^2 \sin^2 \theta (d\phi - N^\phi dt)^2\\ + \frac{\tilde{A} (r,\theta)^2}{N (r, \theta, \psi)}^2(dr^2 + r^2d\theta^2).
        \label{eqn:eq4}
\end{multline}
Here $\tilde{A}$ and $\tilde{B}$ are related to $A$ and $B$ of equation (\ref{eqn:eq1}) by :
\begin{equation}
    \tilde{A} = NA,
    \label{eqn:eq5}
\end{equation}
and 
\begin{equation}
    \tilde{B} = NB.
    \label{eqn:eq6}
\end{equation}
The metric coefficients given in equation (\ref{eqn:eq4}) are the components of the metric tensor w.r.t the co-ordinates $(t, r, \theta, \phi)$, but are expressed as functions of co-ordinates $(t, r, \theta, \psi)$. 

In weak gravitational field limit\cite[]{eric1}, equation (\ref{eqn:eq4}) reduces to the form :
\begin{multline}
        g_{\alpha \beta} dx^\alpha dx^\beta = -[1 + 2\nu (r, \theta, \psi)]dt^2 + [1 - 2\nu(r, \theta, \psi)] \times \\ (dr^2 + r^2d\theta^2 + r^2\in^2\theta d\phi^2).
        \label{eqn:eq7}
\end{multline}
Equation (\ref{eqn:eq7}) represents the well-known metric from which the Einstein equations reduces to the Newtonian equations with gravitational potential $\nu$.

Without any spacetime symmetry or any weak gravitational field assumption, an elliptic equation can be obtained for the gravitational potential ($\nu$) via the 3 + 1 formalism in general relativity \cite[]{eric1}. This is done by taking the trace of the projection of the Einstein equations onto the hyper-surfaces $\sum_t$ defined by $t = $ constant and making use of the Hamiltonian constraint equation \cite[]{cite51}
\begin{equation}
    \nu|_{i}^{i} = 4\pi(E + S_i^i) - \nu^{|i}\nu_{|i} + K_{ij}^{ij} - \frac{1}{N}\bigg(\frac{\partial K}{\partial t} + N^iK_{|i}\bigg).
    \label{eqn:eq8}
\end{equation}
The indices $i,j$ run from 1 to 3, $\nu_{|i}$ is the co-variant derivative of $\nu$ w.r.t the  3-metric induced by $g$ in the hyper-surfaces $\sum_t$, $E$ and $S_{ij}$ are the fluid energy density and stress tensor measured by an Eulerian observer \cite[]{eric1}, $K_{ij}$ is the extrinsic curvature tensor of $\sum_{t}$, $K = K_i^i$ and the $N^i$s are components of the shift vector \cite[]{eric2}. Thus, from equation (\ref{eqn:eq4}) and (\ref{eqn:eq8}), the following expression is obtained.
\begin{multline}
        \bigtriangleup_3 \nu = 4\pi\frac{\tilde{A}^2}{N^2}(E + S_i^i) + \\ \frac{\tilde{B}^2}{2N^4}r^2\sin^2\theta\bigg[\bigg(\frac{\partial N^{\phi}}{\partial r}\bigg)^2 + \frac{1}{r^2}\bigg(\frac{\partial N^{\phi}}{\partial \theta}\bigg)^2\bigg]\\
        - \frac{\partial \nu}{\partial r}\frac{\partial ln \tilde{B}}{\partial r} - \frac{1}{r^2}\frac{\partial \nu}{\partial \theta}\frac{\partial ln \tilde{B}}{\partial r}.
        \label{eqn:eq9}
\end{multline}
Here, $\bigtriangleup_3$ stands for the Laplacian operator in spherical polar co-ordinated in 3D flat space. 
\begin{multline}
        \bigtriangleup_3 := \frac{\partial^2}{\partial r^2} + \frac{2}{r} \frac{\partial}{\partial r} + \frac{1}{r^2}\frac{\partial^2}{\partial \theta^2} \\
        + \frac{1}{r^2\tan \theta}\frac{\partial}{\partial \theta} +  \frac{1}{r^2\sin^2\theta}\frac{\partial}{\partial \psi}.
        \label{eqn:eq10}
\end{multline}
The equations for the remaining gravitational potential terms namely $N^{\phi}, \tilde{A}$ and $\tilde{B}$ are same as that of the axisymmetric case. They have been explicitly derived in \cite{eric_lecture} (see equations $3.14$ to $3.17$) by incorporating a fully three dimensional treatment of the shift vector \cite[]{eric2} . They are  of the form
\begin{equation}
        \tilde{\bigtriangleup}_3(N^{\phi}r\sin \theta) = \sigma_{N\phi},
        \label{eqn:eq11}
\end{equation}
\begin{equation}
        \tilde{\bigtriangleup}_2(\tilde{B}r \sin\theta) = \sigma_{\tilde{B}},
        \label{eqn:eq12}
\end{equation}
\begin{equation}
        \tilde{\bigtriangleup}_2(ln \tilde{A}) = \sigma_{\tilde{A}}.
        \label{eqn:eq13}
\end{equation}
Where,
\begin{equation}
        \bigtriangleup_2 = \frac{\partial ^2}{\partial r^2} + \frac{1}{r} \frac{\partial}{\partial r} + \frac{1}{r^2}\frac{\partial^2}{\partial \theta^2},
        \label{eqn:eq14}
\end{equation}
\begin{equation}
        \tilde{\bigtriangleup}_2 = \frac{\partial ^2}{\partial r^2} + \frac{2}{r} \frac{\partial}{\partial r} + \frac{1}{r^2}\frac{\partial^2}{\partial \theta^2} + \frac{1}{r^2 \tan \theta}\frac{\partial}{\partial \theta} - \frac{1}{r^2 \sin^2 \theta}.
        \label{eqn:eq15}
\end{equation}
The R.H.S of the equations (\ref{eqn:eq11}) - (\ref{eqn:eq13}) i.e. the sources $\sigma$ might contain some non-axisymmetric terms such as the matter density, pressure or the potential ($\nu$). The average of these terms over $\psi$ is taken so that the solutions of $N^{\phi}$, $\tilde{A}$ and $\tilde{B}$ remain axisymmetric and consistent with the approximation taken previously.

\section{Lorene/ Nrotstar Numerical Code}
\label{sec:Lorene}
The Nrotstar code is a free software based on the C++ LORENE\footnote{http://www.lorene.obspm.fr/} library. It is a descendant of rotstar, a previous LORENE code described in many astrophysical studies. The equations (\ref{eqn:eq9}) and (\ref{eqn:eq11}) - (\ref{eqn:eq13})are solved by an iterative scheme, starting from very rudimentary initial conditions; a spherically symmetric star with constant matter density and a flat metric tensor\cite[]{eric_lecture, eric1, eric2}.
From, the given density and pressure profiles, the Poisson type gravitational equations (\ref{eqn:eq9}) and (\ref{eqn:eq11}) - (\ref{eqn:eq13}) are solved to obtain new values of $\nu, N^{\phi}, \tilde{A}$ and $\tilde{B}$. The value of $\nu$ obtained is now put into the equation \footnote{This equation is known as the first integral of motion \cite[]{eric1}.}. 
\begin{equation}
    H + \nu - \ln \Gamma = constant. 
\end{equation}
\label{eqn16}
where H is the log enthalpy\cite[]{eric1}, $\nu $ is the gravitational potential as defined in section (\ref{sec:Theoretical Model}) and $\Gamma$ is the Lorentz factor between a Eulerian observer and a fluid co-moving observer.

This yields a new value of log enthalpy $H$. Once this has been done, new density and pressure profiles are computed from the equation of state and a new iteration is followed thereafter \footnote{This procedure is also referred to as the self-consistent field method\cite[]{cite53}.}.
For the first few steps of the procedure, $\Omega$ is set to zero such that spherical symmetry of the solution is not broken. At a certain step, the rotation is switched on i.e. $\Omega$ is no longer zero, and consequently, the solution is now axisymmetric. The solutions converge to a certain state with the increase in the no. of iterations, and this state represents a stationary axisymmetric solution. After this, at a certain step $J_0$, a perturbative term is added to the potential $\nu$   \cite[]{eric2}.
\begin{equation}
    \delta \nu = -\epsilon H_c(r \sin \theta \cos \psi)^2.
    \label{eqn:eq17}
\end{equation}
$H_c$ is the log enthalpy at the center of the star and $\epsilon$ is a constant of the order $10^{-6}$. The form given in equation \ref{eqn:eq17} excites the bar mode $l = 2, m = \pm2$ \cite[]{eric1, eric2, cite3, citeJ}. According to the first integral of motion, the enthalpy $H$ gets modified by $-\delta \nu$ and becomes non-axisymmetric along with the matter density and pressure following the EOS. For steps $J > J_0 + 1$, the perturbation is switched off, but the solution tends to remain non-axisymmetric owing to the non-axisymmetric parts of the matter density, the pressure and $N$ which is present in the source of the three-dimensional Poisson equation (\ref{eqn:eq10}) for the potential $\nu$\cite[]{eric1, eric2}.

After the perturbative term has been added, at each step, the quantity $q$ \cite[]{cite55, citeJ} is evaluated, which is defined as:
\begin{equation}
    q = \max \mid \hat{\nu}_{2}\mid.
\end{equation}
where $\hat{\nu}_{2}$ is the $m=2$ coefficient in the Fourier expansion of the $\psi$ part of gravitational potential $\nu$. This quantity is used to study the evolution of triaxial perturbation introduced in equation (\ref{eqn:eq17}). The axisymmetric configuration is stable if the $q$ decays and tends to zero as the iteration proceeds. However, in case of marginally stable configurations, the decay or the growth of perturbation turns out to be pretty small. Hence, it useful to useful to use the relative growth rate of $q$ which has been defined in equation (\ref{eq_new1})  for the $j^{th}$ step in the iteration \cite[]{citeJ,cite55}.
\begin{equation}
\dot{q_j} = \frac{q_j - q_{j-1}}{q_{j-1}}.
    \label{eq_new1}
\end{equation}
Here $j-1$ refers to the previous step in the iteration while $\dot{q} $ is defined as the growth rate in triaxial perturbation and used to determine the stability of axisymmetric configurations as seen in subsequent sections of this article.

\section{Results for Polytropes}
\label{sec:polytropes}

\subsection{Newtonian Polytropes}
\label{sec:NPolytropes}
Polytropic equations of state represent simplified but consistent extensions of realistic equations of state.  Triaxial instabilities of such stars are investigated. Once, the triaxial perturbation is introduced as described in section (\ref{sec:Lorene}), its behavior is studied with the variation of the frequency of rotation. An important parameter in determining the stability of the star with respect to triaxial perturbation is the Growth rate of triaxial perturbation as mentioned in section (\ref{sec:Lorene}).
If this quantity turns out to be positive at a particular frequency of rotation, it can be inferred that symmetry breaking occurs or triaxial instability sets in at that point. On the other hand, if this quantity turns out to be negative, it would mean that the final solution is axisymmetric and the star doesn't break its symmetry.

\begin{figure}
\centering
\includegraphics[width = 0.48\textwidth]{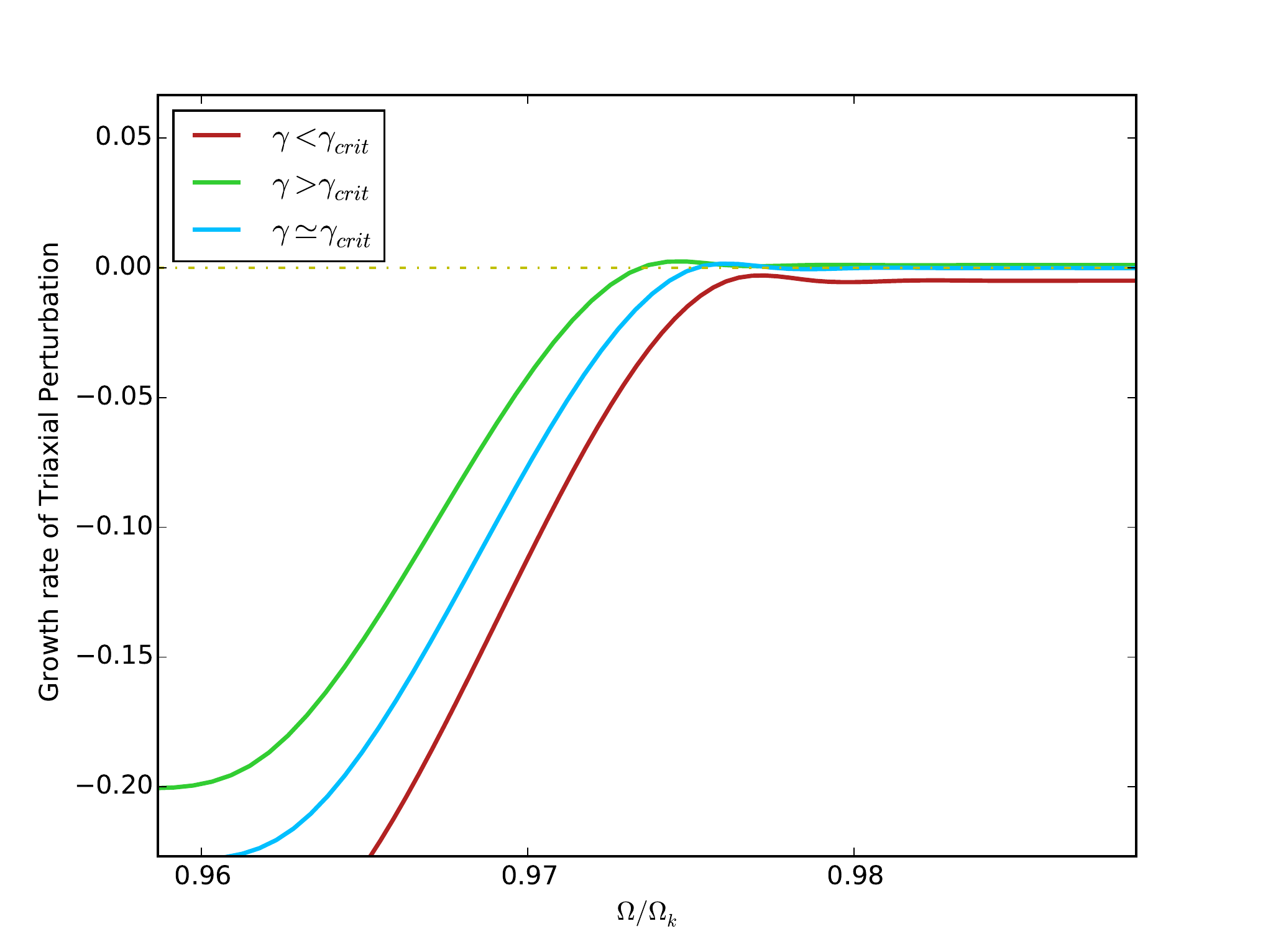}
\caption{Plot showing the variation of growth rate of triaxial perturbation with respect to normalized rotation frequency $( \Omega/\Omega_{k})$ at $1.4 M\odot$ for Newtonian polytropes with different values of $\gamma$.}
\label{fig:newtonian1}
\end{figure}

Figure (\ref{fig:newtonian1}) illustrates the triaxial stability of three different stars with polytropic equations of state. The normalized frequency for a particular star is determined by taking the ratio of the frequency of rotation to the Keplerian frequency (mass-shedding limit) for that particular configuration $( \Omega/\Omega_{k})$. The plots have been obtained by keeping the Baryon mass of the star constant for all the values Rotation frequency ($\Omega$). This is achieved by varying the value of central enthalpy ($H_c$) \cite[]{eric1} in accordance with the rotation frequency so as to keep the baryon mass constant at $1.4 M_{\odot}.$

It has been observed that as adiabatic index ($\gamma$) increases and approaches the value of $\gamma_{crit}$ \footnote{The value of $\gamma_{crit}$ is estimated at $2.238 \pm 0.002$ \cite[]{eric1,cite34}.}, the values of growth rate of triaxial perturbation becomes positive and thus the instability
sets in at a lower value of normalized frequency. This result is pretty much consistent with \cite{eric1} because as the value of $\gamma$ increases, the stiffness of the EOS increases. Hence, the axisymmetry is broken at a lower value of $\Omega/\Omega_{k}$. Other tests to examine the validity of the Axisymmetric part of the code have been explicitly performed in \cite{eric2, eric1}. These include an estimation of $\gamma_{crit}$ and $T/W_{crit}$.

\subsection{Relativistic Polytropes}
\label{sec:Relativistic_polytropes}
Relativistic polytropes represent a natural extension of the classical results. Also, thermodynamic inconsistencies, which is prevalent among tabulated EOS is not present in this case. Thus, it provides an approximate but consistent model for real stars thereby allowing investigations of relativistic effects.

\begin{figure}
\centering
\includegraphics[width = 0.48\textwidth]{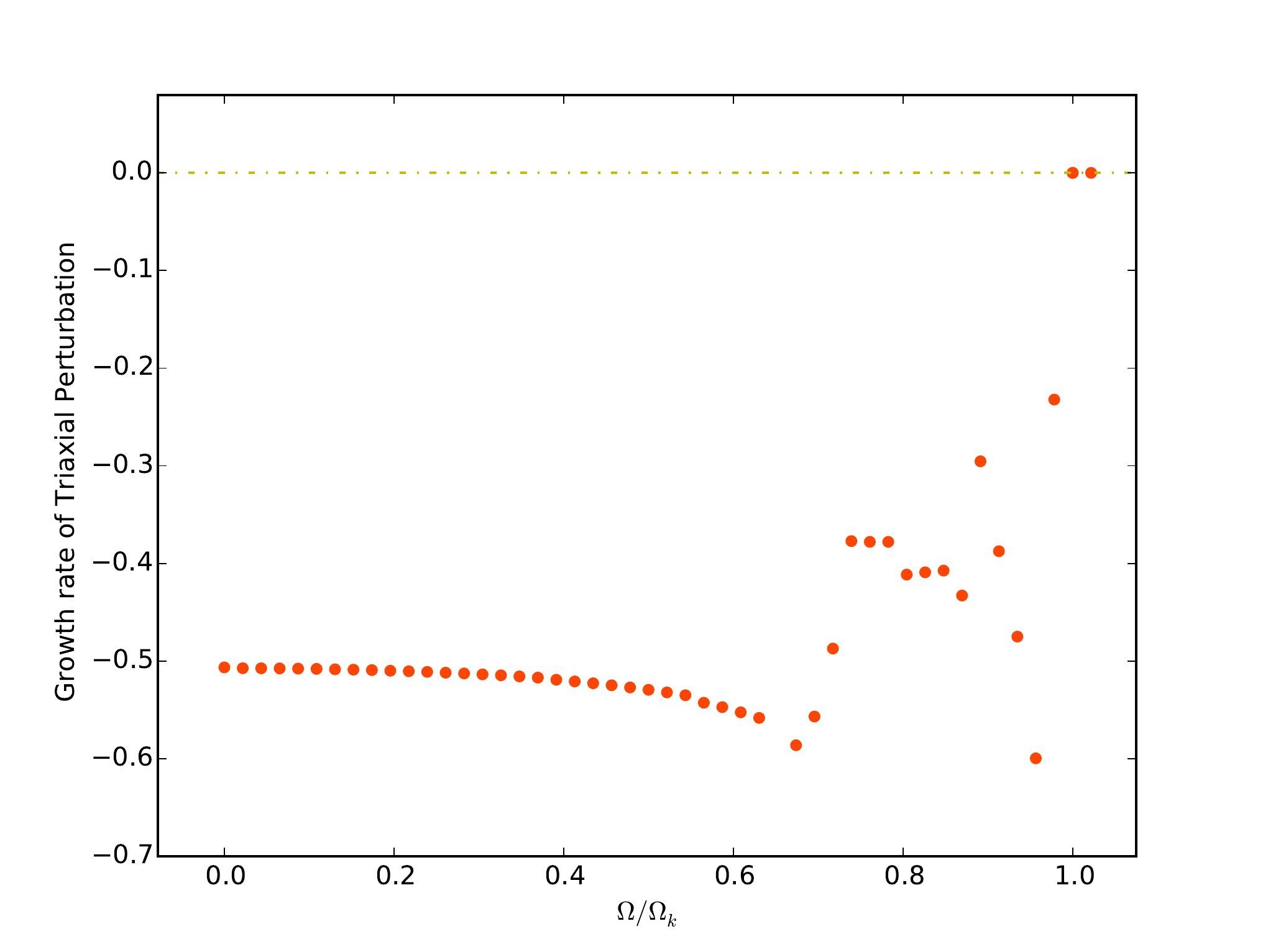}
\caption{Plot showing the unpredictable behaviour of the solutions at frequencies greater than 0.6 $\Omega/\Omega_{k}$ for a Relativistic Polytrope at $M_b = 1.4 M\odot$.}
\label{fig:R_polytrope}
\end{figure}

The same routine as mentioned in section (\ref{sec:NPolytropes}) has been performed for relativistic polytropes i.e. investigating the growth rate of triaxial perturbation with respect to normalized rotation frequency for a constant value of Baryon Mass ($1.4 M_{\odot}$). The results obtained for this case weren't exactly similar to that of the Newtonian case. Figure (\ref{fig:R_polytrope}) shows that the solutions to the field equations did not converge for all values of rotation frequencies especially at values greater than 0.6 $\Omega/\Omega_{k}$ giving out large values of relative errors on Virial theorem (GRV2 and GRV3)\footnote{The quantities GRV2 and GRV3 have been explicitly defined in equations ($2.27$) and ($2.28$) in \cite{cite55}.} and unpredictable values of growth rate of triaxial perturbation thereby leading to ambiguous conclusions with regard to the stability of the star.

This ambiguity was resolved by varying a few numerical parameters to obtain convergence of the solution at higher frequencies. The parameters that were tweaked include: 

\begin{itemize}
    \item The number of grid points in both $\theta$ and $\phi$ domain \footnote{It has been established in \cite{eric1} that relative errors on Virial theorem sometimes depend on this parameter and can be brought within the acceptable limit ($\simeq 10^{-5}$ for polytropic EOSs) by increasing the no of grid points in each domain.}.
    \item Amplitude of Triaxial Perturbation (the constant $\epsilon$ in equation (\ref{eqn:eq17}).
    \item Step at which the triaxial perturbation is switched on ($J_0$).
    \item The total number of steps in the computation.
    \item Relaxation factor $\vee$ in the main iteration\footnote{The value of central enthalpy gets updated after each iteration following the equation $ H^J = \vee H^J + (1 - \vee)H^{J -1}$ where $J$ refers to the current iteration step and $\vee$ refers to the relaxation factor.}.
    \item Threshold on ($dH/dr_{eq})/dH/dr_{pole}$) \footnote{Here $H$ is the log enthalpy, $r_{eq}$ is the equatorial radius while $r_{pole}$ is the polar radius of the star. } for the mapping adaptation.
\end{itemize}

\begin{figure}
\centering
\includegraphics[width = 0.48\textwidth]{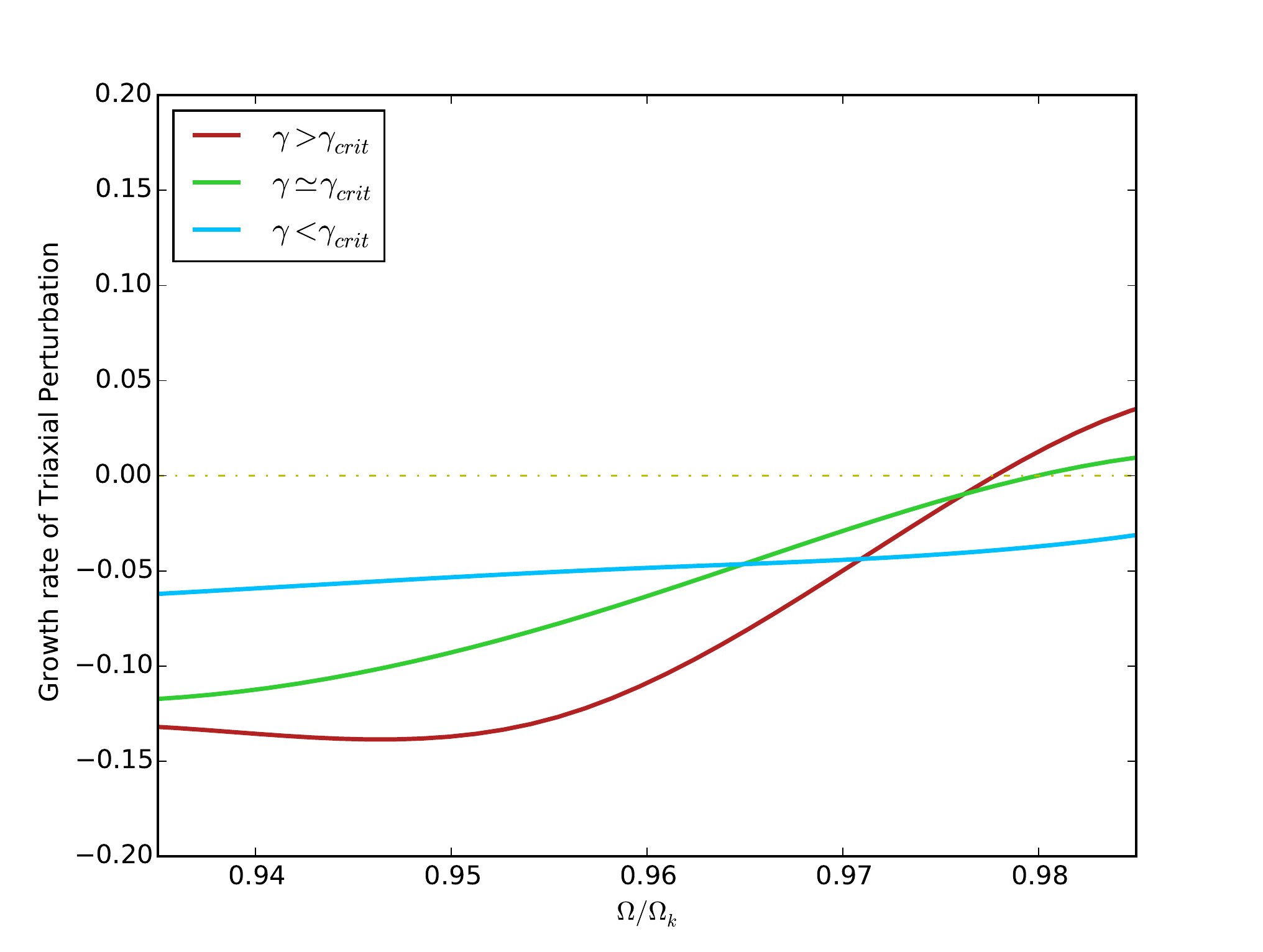}
\caption{Plot showing the variation of Growth rate of triaxial perturbation with respect to normalized rotation frequency  $( \Omega/\Omega_{k})$ at $1.4 M\odot$ for relativistic polytropes for different values of $\gamma$. }
\label{fig:R_polytrope1}
\end{figure}

An optimal combination of the variation of the mentioned parameters led to the convergence of the solutions at higher frequencies and lower errors for GRV2 and GRV3 were obtained for frequencies greater than 0.6 $\Omega/\Omega_{k}$ as seen in Figure (\ref{fig:R_polytrope1}). 

From, figure (\ref{fig:R_polytrope1}), it can again be observed that the triaxial instability sets in at lower values of $\Omega/\Omega_{k}$ for high values of $\gamma$. Thus, the results obtained are consistent with the ones obtained in section (\ref{sec:NPolytropes}).
Once, inferences regarding triaxial instabilities for polytropes were obtained, the same methodology would be utilized for checking the instabilities for realistic equations of state. To check if this method of variation of parameters is sufficient to converge the solution at all values of rotation frequency, a tabulated EOS was created by using the polytropic equation described below \cite[]{cite58}. 

\begin{equation}
    p = K\rho^{\gamma}.
\label{eqn:eq18}
\end{equation}

Here $p$ is the pressure, $K$ is the pressure constant, and $\rho$ is the matter density. The same routine was performed for $\gamma = 2.3$ and $K = 0.04$. As $\gamma > \gamma_{crit}$, the Growth rate of triaxial perturbation becomes positive at $\Omega < \Omega_k$.

\begin{figure}
\centering
\includegraphics[width = 0.48\textwidth]{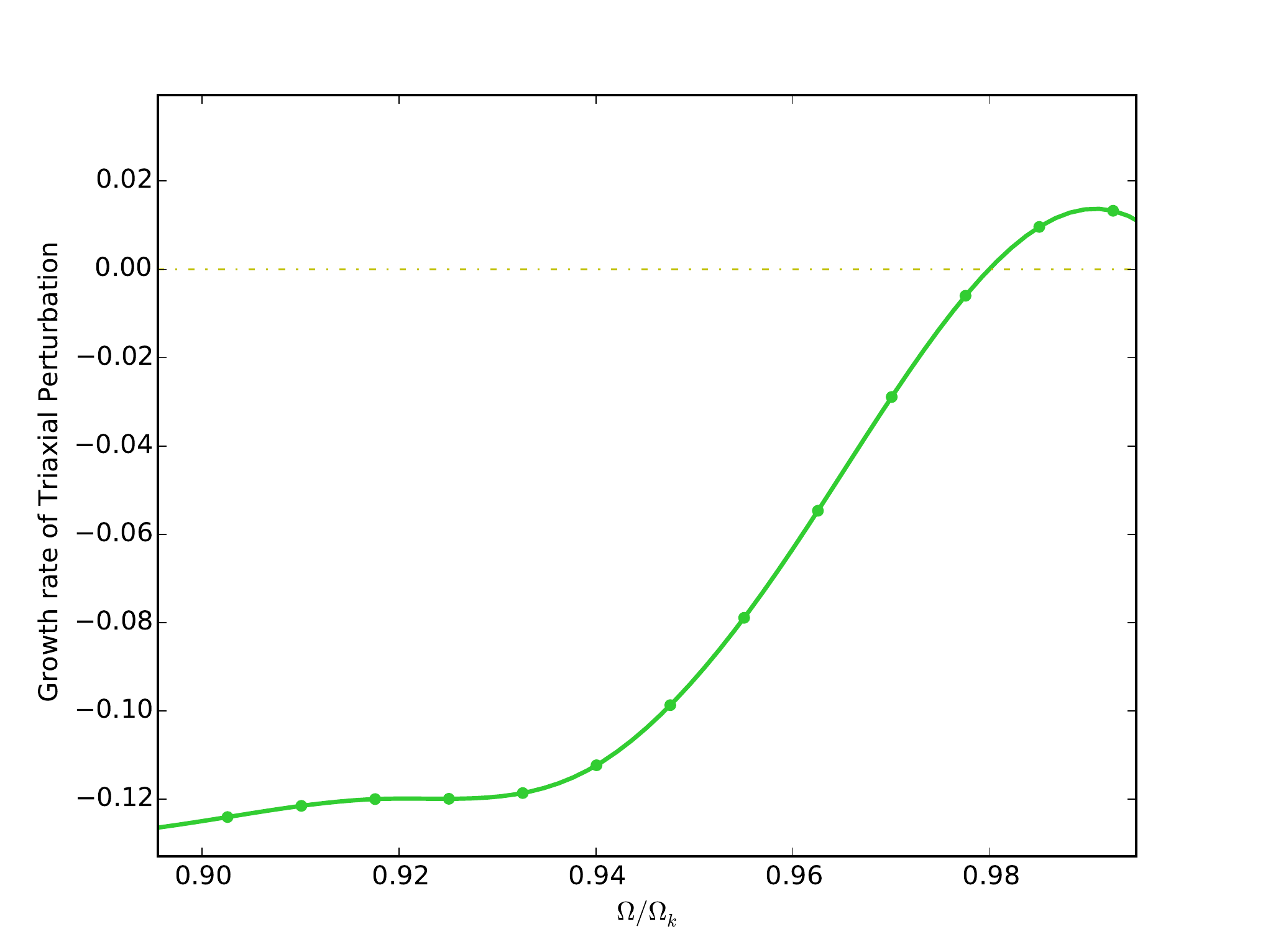}
\caption{Growth rate of Triaxial Perturbation vs $\Omega/\Omega_{k}$ on using a tabulated version of a polytropic EOS. for $\gamma = 2.3$ and $M_b = 1.4 M\odot$.}
\label{fig:R_polytrope2}
\end{figure}

From, figure (\ref{fig:R_polytrope2}), it is observed that the
solutions converge for all
values of rotation frequency $\Omega$ and the results are similar to that of the Newtonian case.

\section{Results for Realistic Equations of state}
\label{sec:EOS}
Tests on the capability of the code to determine triaxial instabilities for Realistic EOSs \cite[]{cite26, cite15, cite16, cite20, cite4, cite5, cite22, cite23} have been performed in appendix (\ref{code_test}) and the method that has been followed to reduce the Virial errors (GRV2 and GRV3) have been elaborated in appendix (\ref{GRV_errors}). Triaxial instabilities were then determined on various realistic equations of state available on the CompOSE \footnote{http://compose.obspm.fr/} database. The method for investigating triaxial instability is similar to the one
mentioned in the previous sections i.e observe the variation in the Growth rate of triaxial perturbation with respect to the change in frequency for a constant value of Baryon Mass.

\begin{figure}
\centering
\includegraphics[width = 0.48\textwidth]{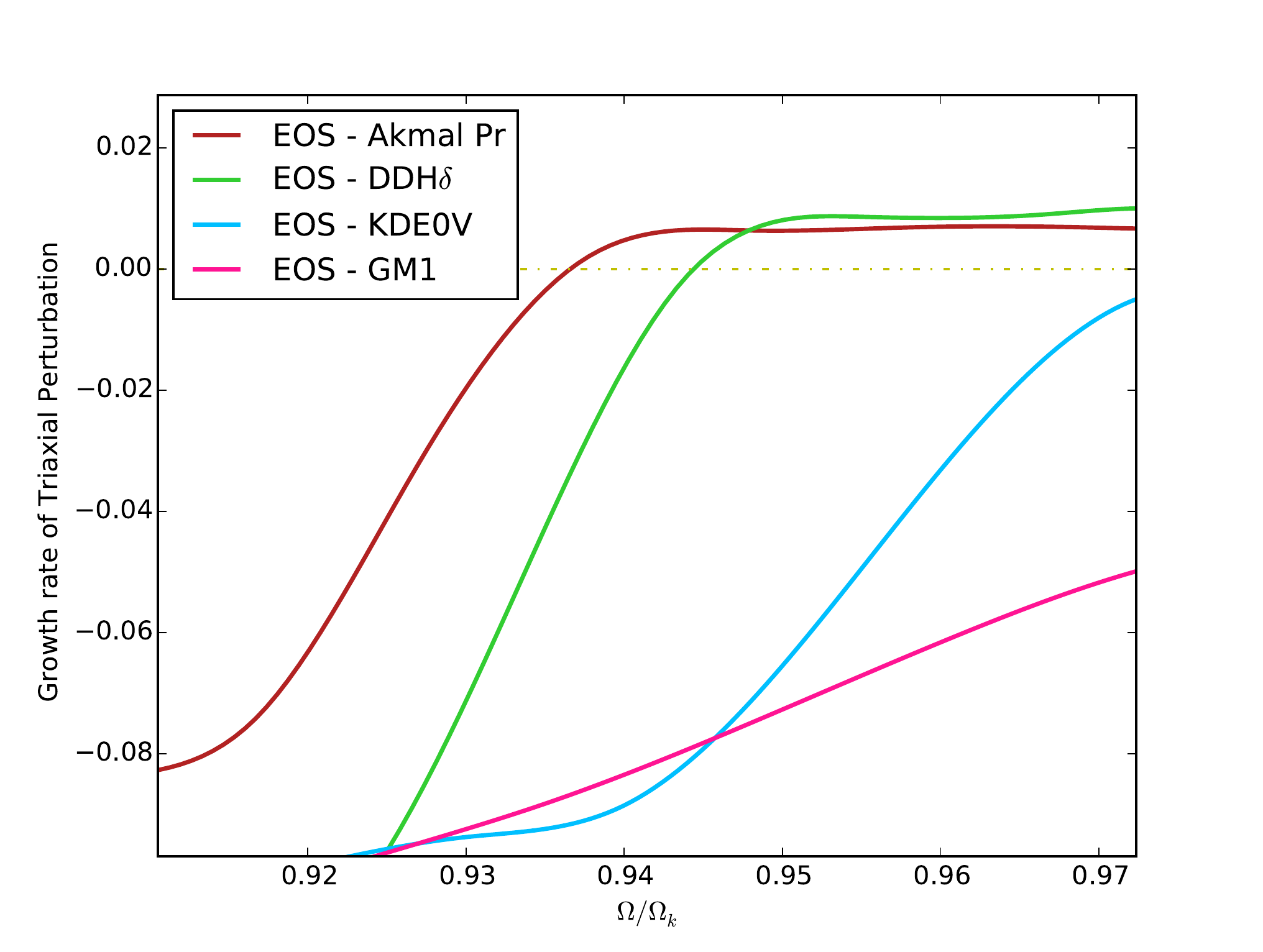}
\caption{Plot of Growth rate of Triaxial Perturbation vs normalized frequency $( \Omega/\Omega_{k})$ for EOS AkmalPr, KDE0V, DDh$\delta$, GM1  at $M_b = 1.4 M_{\odot}$.}
\label{fig:EOSs}
\end{figure}

From Figure (\ref{fig:EOSs}), it can be inferred that for EOS KDE0V and EOS GM1, no symmetry breaking occurs before the rotation frequency ($\Omega$) reaches the Kepler frequency ($\Omega_k$) while for EOS Akmal-pr and EOS DDH$\delta$, the Growth rate of triaxial perturbation becomes positive. Hence, symmetry breaking occurs at $M_b = 1.4 M_{\odot}$ . 

\begin{figure}
\centering
\includegraphics[width = 0.48\textwidth]{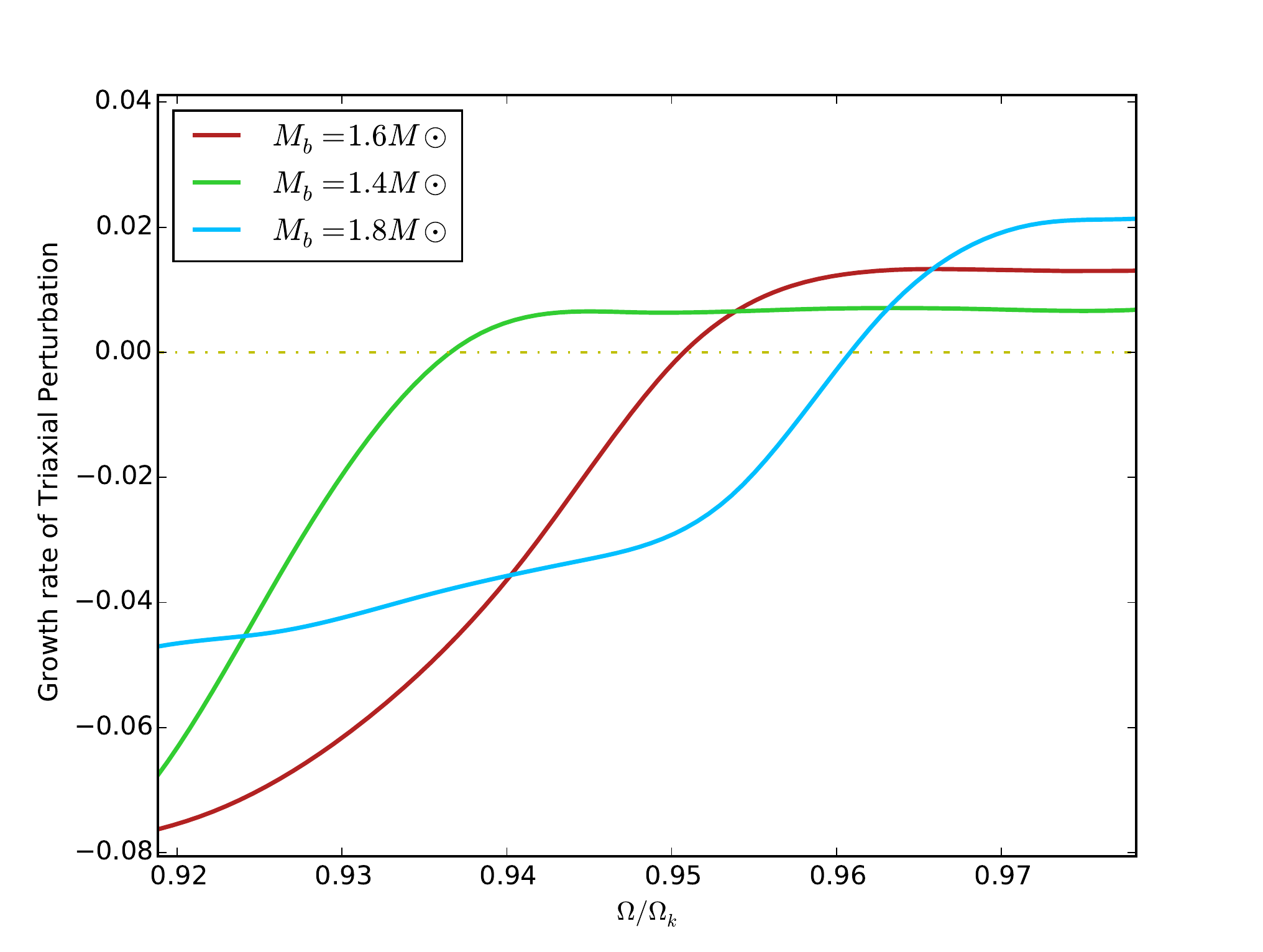}
\caption{Plot of Growth rate of Triaxial Perturbation vs normalized frequency $( \Omega/\Omega_{k})$ for EOS AkmalPr at $1.4, 1.6 $ and $1.8 M_{\odot}$.}
\label{fig:akmal_comp1}
\end{figure}

\begin{figure}
\centering
\includegraphics[width = 0.48\textwidth]{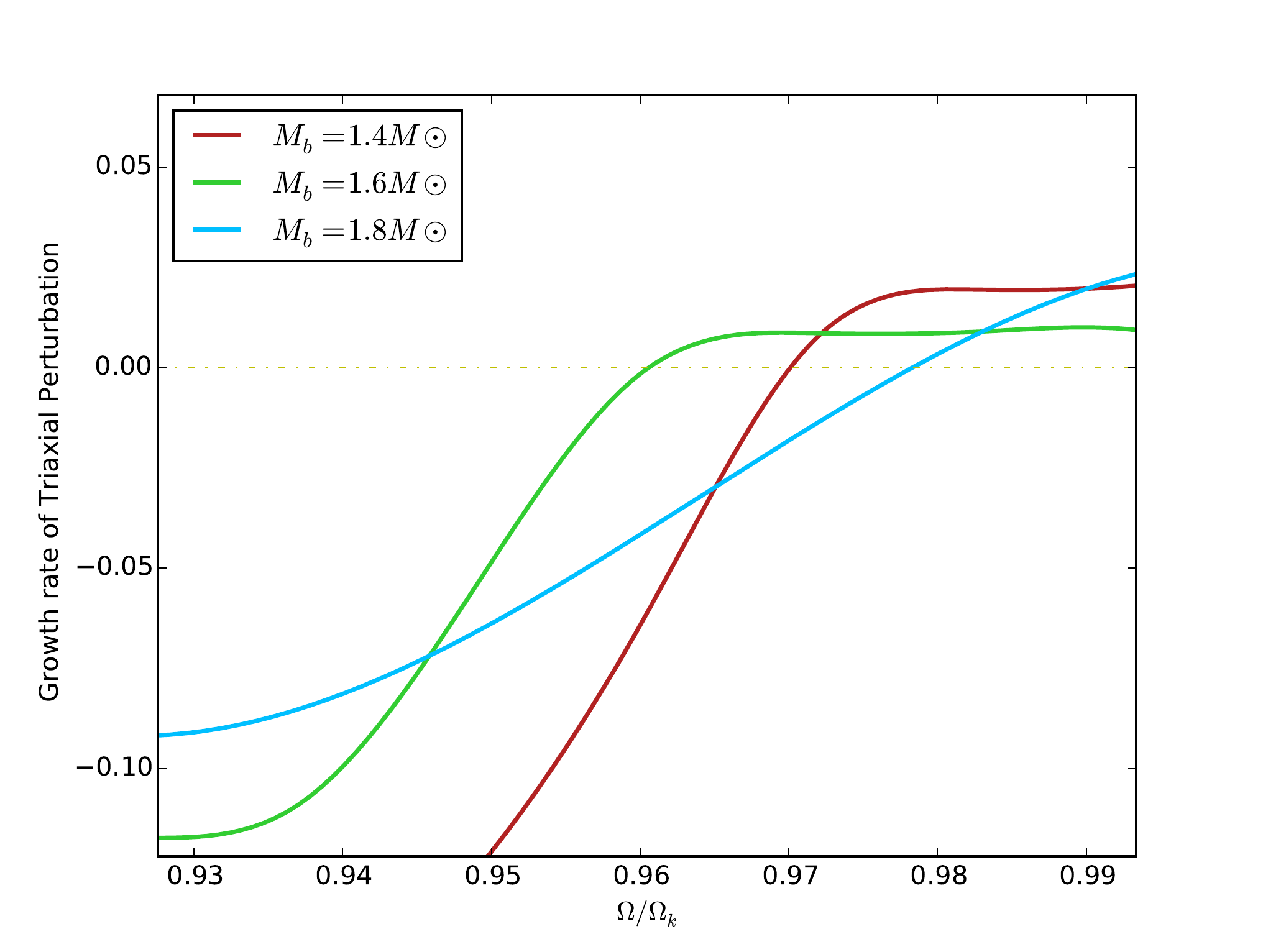}
\caption{Plot of Growth rate of Triaxial Perturbation vs normalized frequency $( \Omega/\Omega_{k})$ for EOS DDH $\delta$ at $1.4, 1.6 $ and $1.8 M_{\odot}$.}
\label{fig:DDH_comp1}
\end{figure}

\begin{figure}
\centering
\includegraphics[width = 0.48\textwidth]{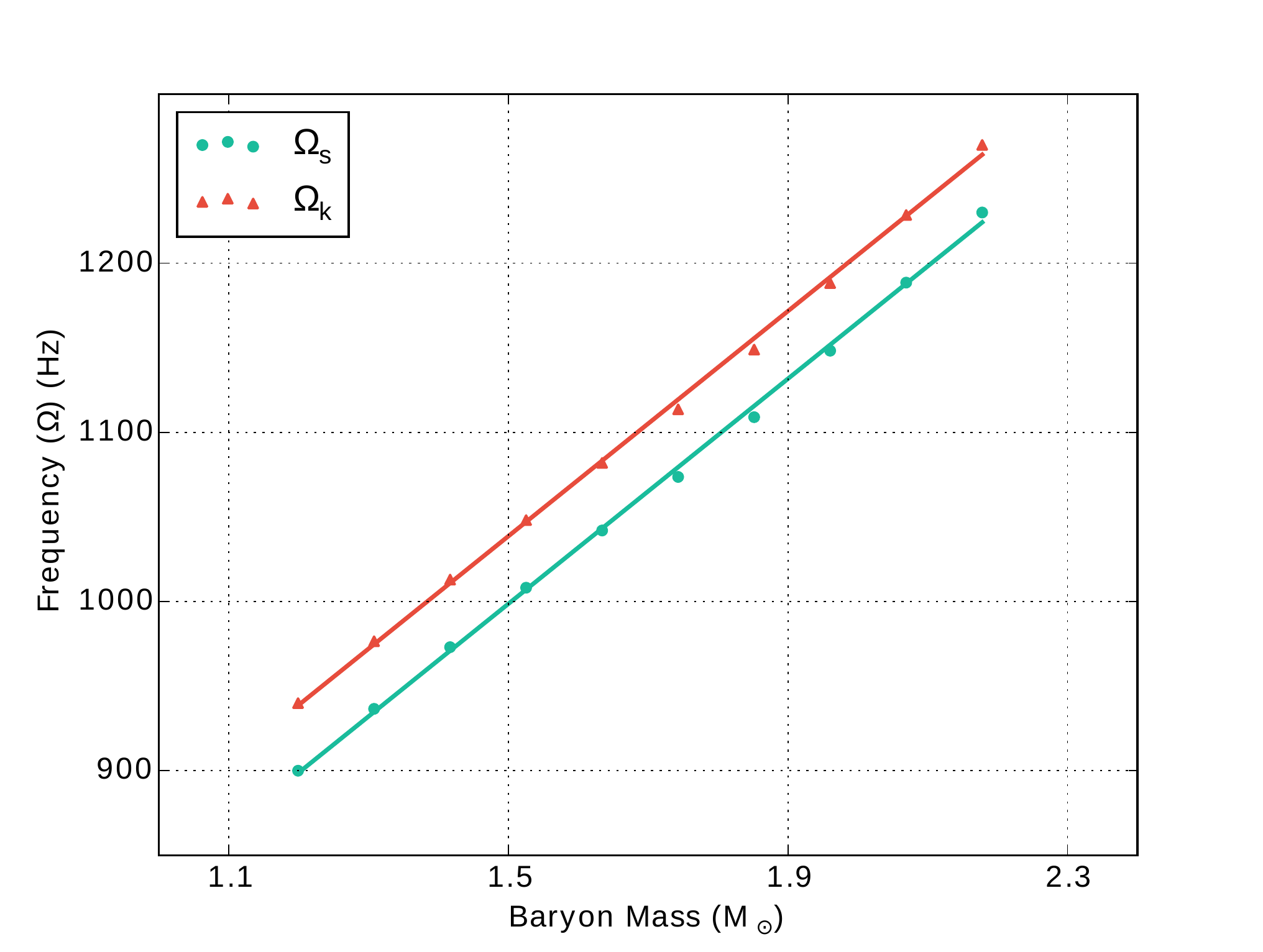}
\caption{Plot showing the variation of $\Omega_s$ and $\Omega_k$ for different values of Baryon Mass $(M_b)$ for EOS Akmal Pr.}
\label{fig:akmal_comp3}
\end{figure}

\begin{figure}
\centering
\includegraphics[width = 0.48\textwidth]{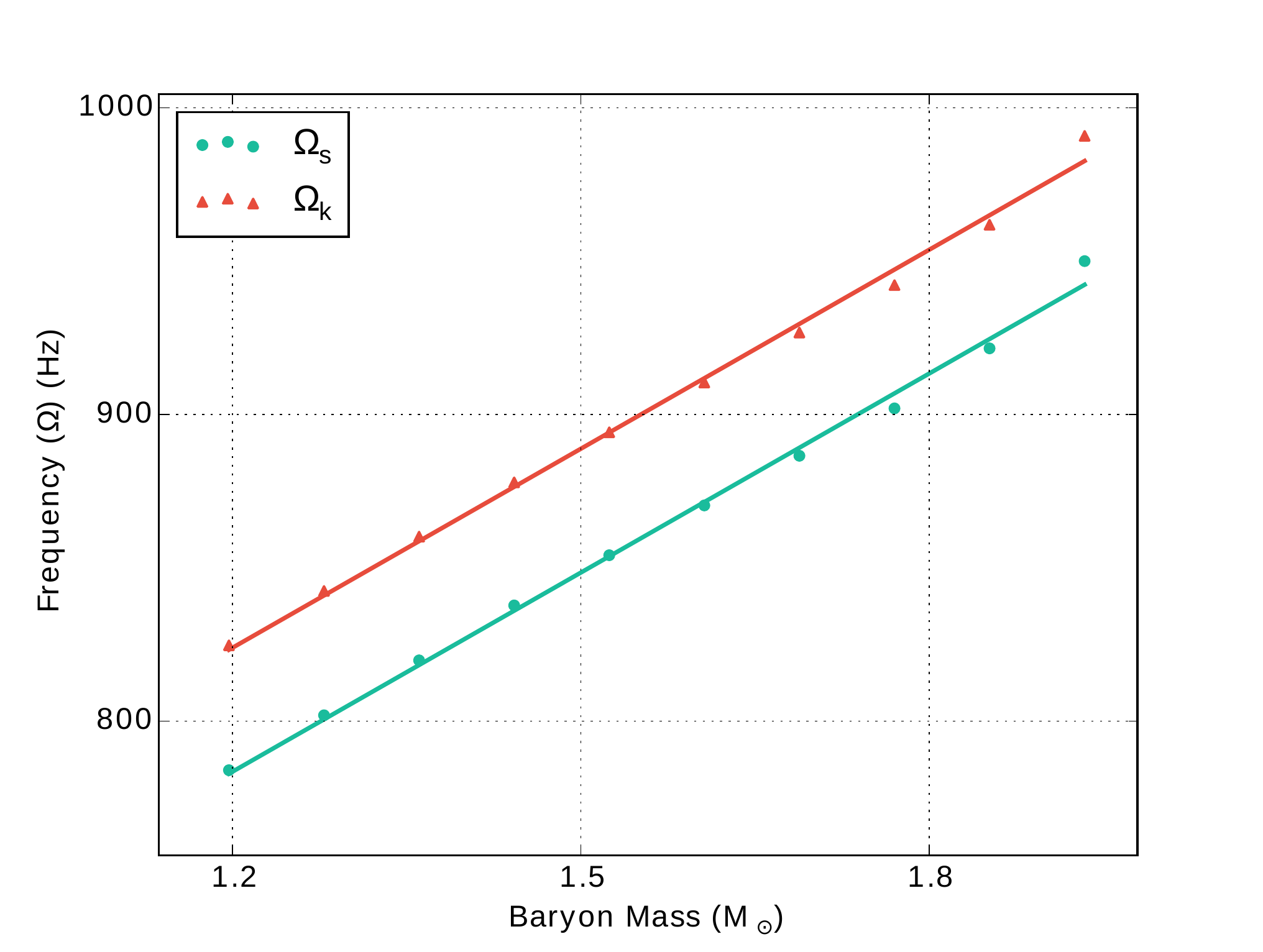}
\caption{Plot showing the variation of $\Omega_s$ and $\Omega_k$ for different values of Baryon Mass $(M_b)$ for EOS DDH$\delta$.}
\label{fig:DDH_comp3}
\end{figure}

The same sequence of operations are done for a higher value of Baryon Mass \cite[]{cite17} the triaxial instability sets in at a higher value of $\Omega/\Omega_{k}$ as shown in Figure (\ref{fig:akmal_comp1}) for EOS Akmal-pr and in Figure (\ref{fig:DDH_comp1}) for EOS-DDH $\delta$. Figures (\ref{fig:akmal_comp3}) and (\ref{fig:DDH_comp3}) explicitly show that the breaking frequency ($\Omega$), the Kepler frequency $(\Omega_k)$, and hence the ratio $\Omega_s/\Omega_k$ increases with an increase in Baryon Mass $(M_b)$. 

\begin{figure}
\centering
\includegraphics[width = 0.48\textwidth]{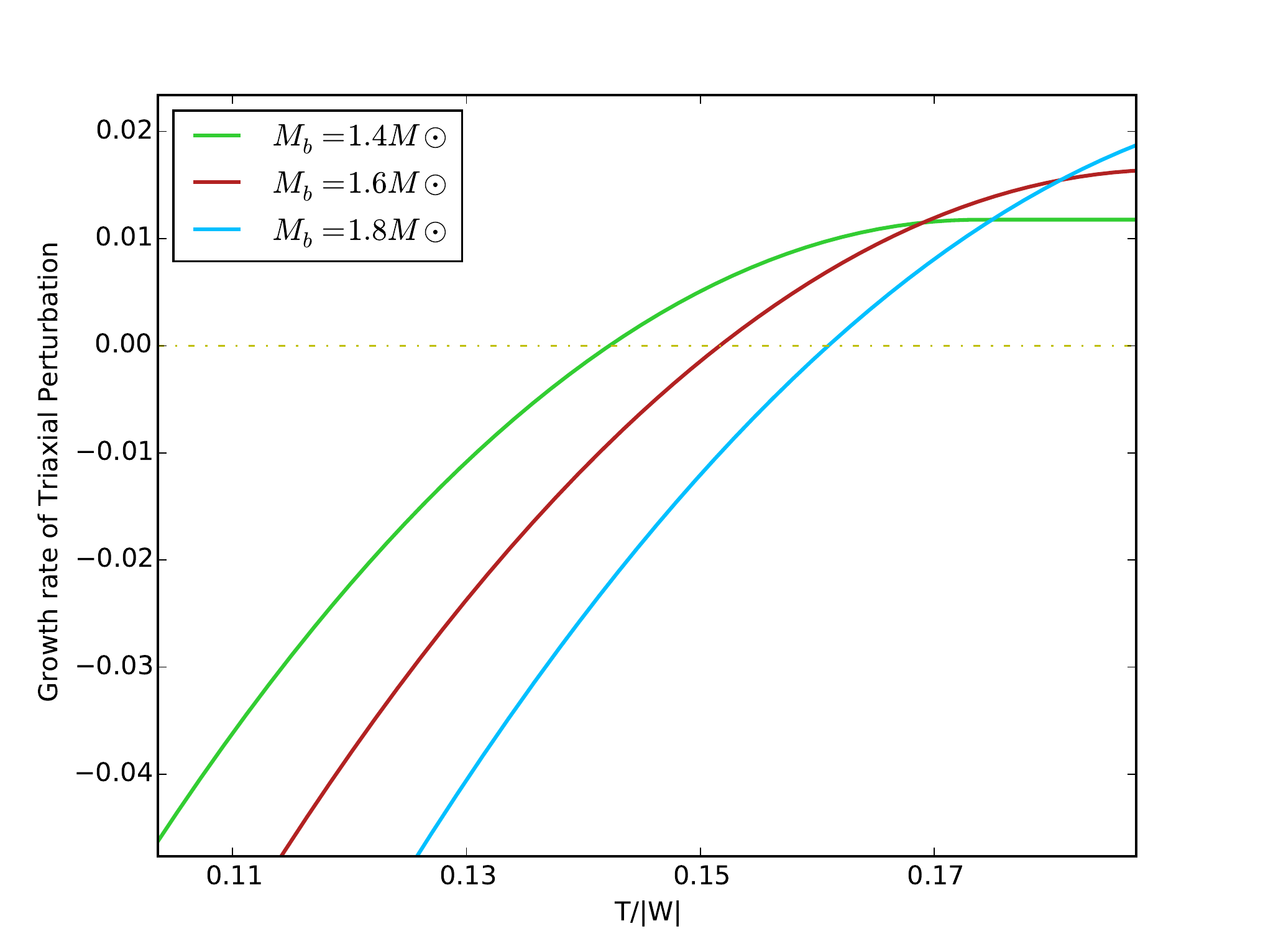}
\caption{Plot of Growth rate of Triaxial Perturbation vs ratio of Kinetic energy to absolute value of Gravitational potential energy $T/|W|$ for EOS Akmal-Pr at $1.4, 1.6 $ and $1.8 M_{\odot}$.}
\label{fig:akmal6}
\end{figure}

\begin{figure}
\centering
\includegraphics[width = 0.48\textwidth]{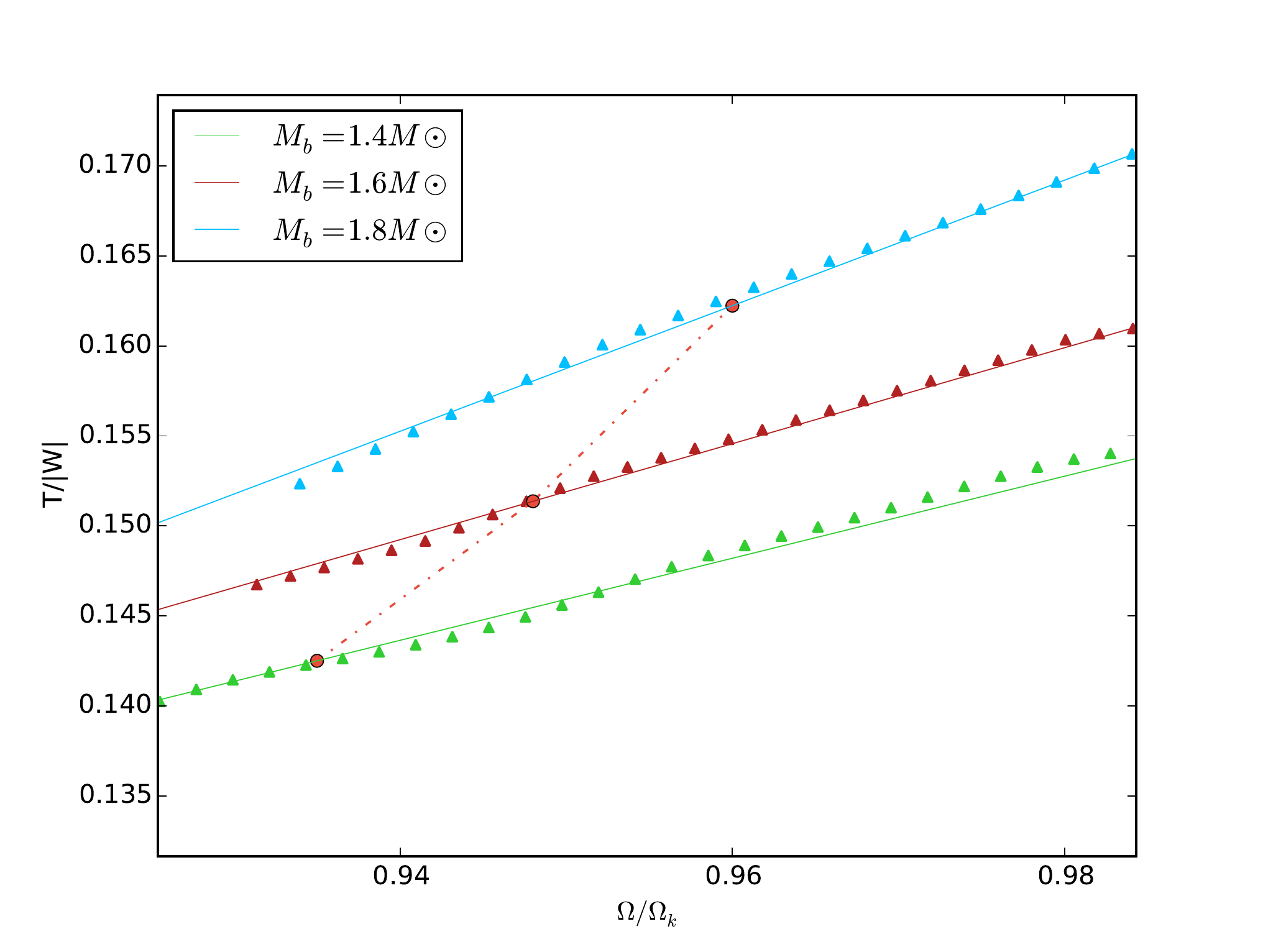}
\caption{Plot of ratio of Kinetic energy to absolute value of Gravitational potential energy $T/|W|$ vs normalized frequency $( \Omega/\Omega_{k})$ for EOS Akmal-Pr at $1.4, 1.6 $ and $1.8 M_{\odot}$. The dotted line connects the symmetry breaking points.}
\label{fig:akmal5}
\end{figure}

\begin{figure}
\centering
\includegraphics[width = 0.48\textwidth]{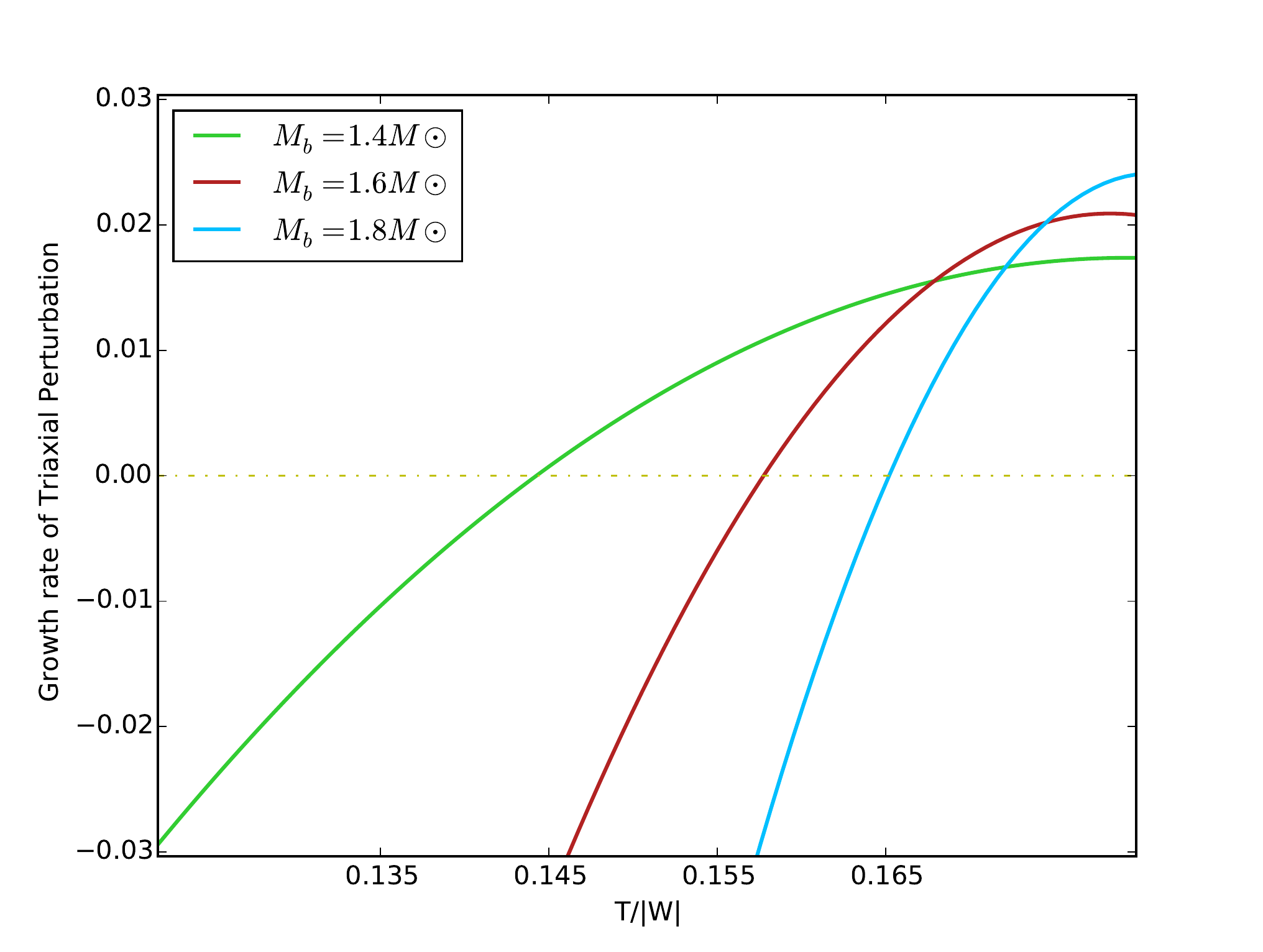}
\caption{Plot of Growth rate of Triaxial Perturbation vs ratio of Kinetic energy to absolute value of Gravitational potential energy $T/|W|$ for EOS DDH $\delta$ at $1.4, 1.6 $ and $1.8 M_{\odot}$.}
\label{fig:DDH6}
\end{figure}

\begin{figure}
\centering
\includegraphics[width = 0.48\textwidth]{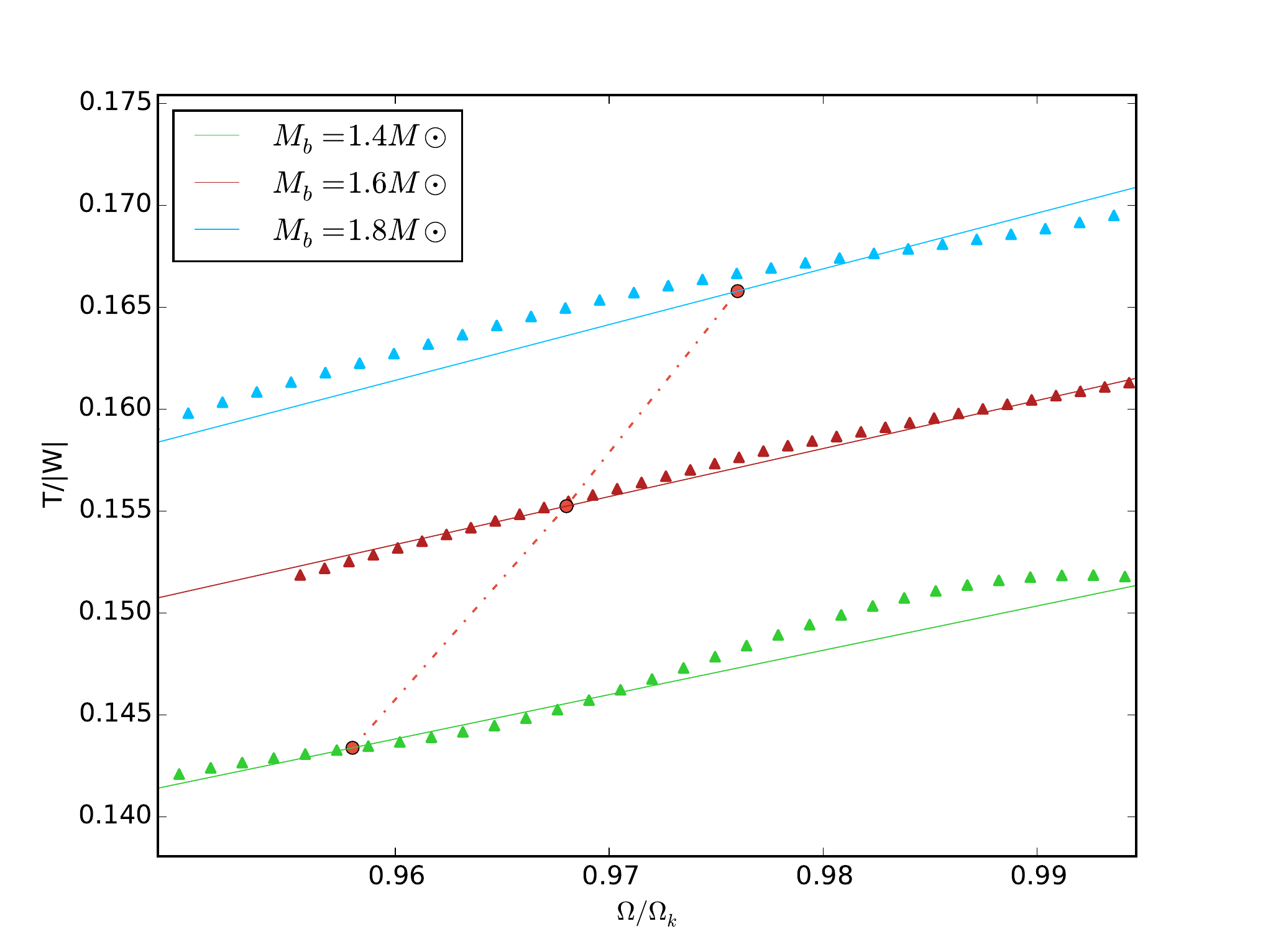}
\caption{Plot of ratio of Kinetic energy to absolute value of Gravitational potential energy $T/|W|$ vs normalized frequency $( \Omega/\Omega_{k})$ for EOS DDH $\delta$ at $1.4, 1.6 $ and $1.8 M_{\odot}$. The dotted line connects the symmetry breaking points.}
\label{fig:DDH5}
\end{figure}

The kinetic energy and hence the ratio of Kinetic energy to absolute value of Gravitational potential energy $T/|W|$ increases with an increase in the rotation frequency of the star. Since, normalized frequency $(\Omega/\Omega_{k})$ at the symmetry breaking point is higher for high mass stars, it would mean that the value of $(T/|W|)_{crit}$ should increase as one proceeds from $M_b = 1.4 M_{\odot}$ to $M_b = 1.8 M_{\odot}$. This behavior has been visualized in figures (\ref{fig:akmal6}) and (\ref{fig:akmal5}) for EOS Akmal-pr and in figures (\ref{fig:DDH6}) and (\ref{fig:DDH5}) for EOS DDH $\delta$.

The increase in ($\Omega_s/\Omega_k$) for higher mass stars can be attributed to the fact that as the mass of the star increases, it becomes more compact i.e. the ratio of the Gravitational mass to the circumferential radius of the star $(M/R_{circ})$ increases. As, the compactness of the star increases, $\Omega_s$ increases, so does $\Omega_k$ and thus the ratio $\Omega_s/\Omega_k$. This hypothesis could be put to test if $\Omega_s$ and $\Omega_k$ are plotted for different values of compactness ($M_g/R_{circ}$).

\begin{figure}
\centering
\includegraphics[width = 0.48\textwidth]{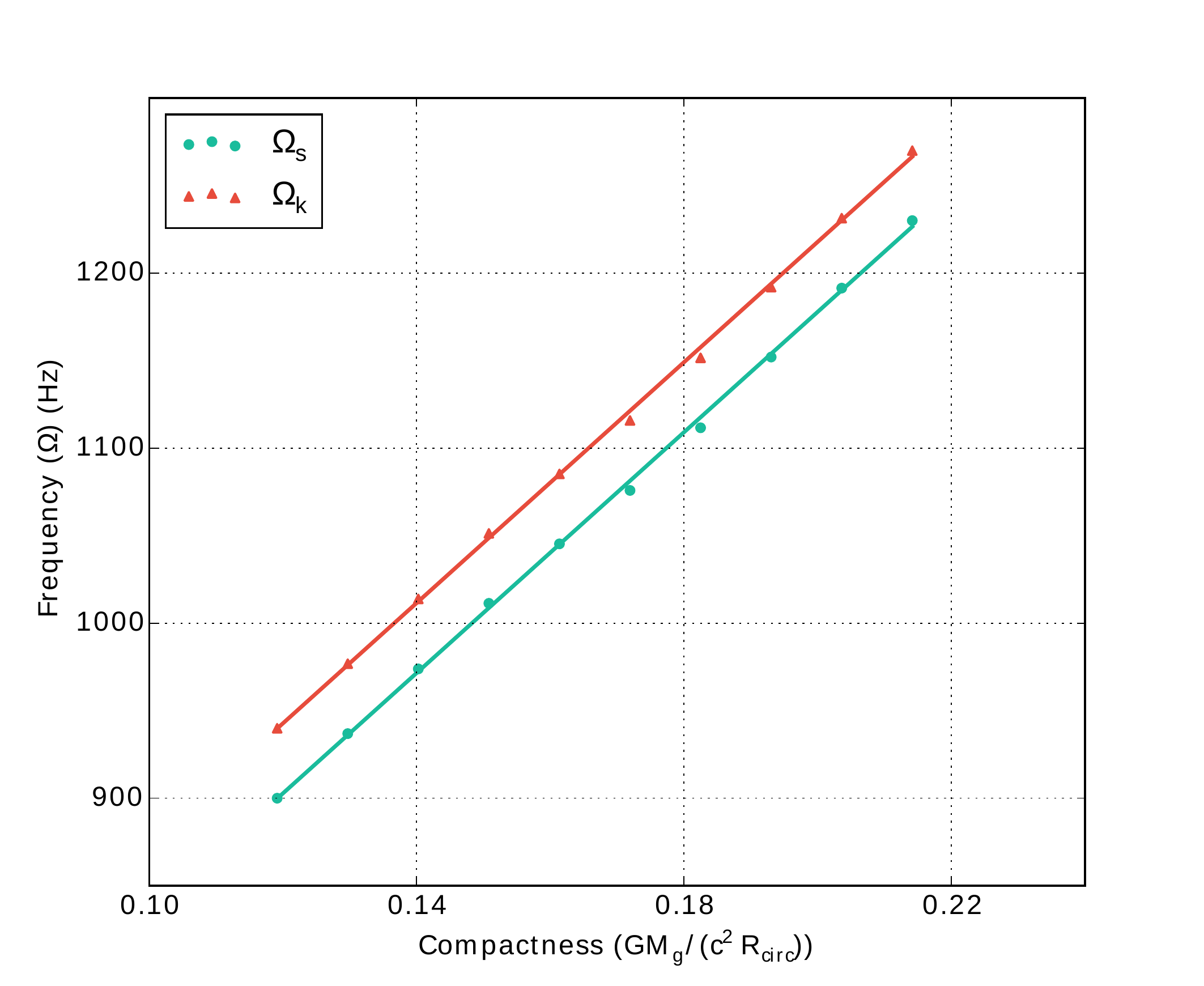}
\caption{Plot showing the variation of $\Omega_s$ and $\Omega_k$ for different values of compactness for EOS Akmal Pr.}
\label{fig:akmal_comp2}
\end{figure}

\begin{figure}
\centering
\includegraphics[width = 0.48\textwidth]{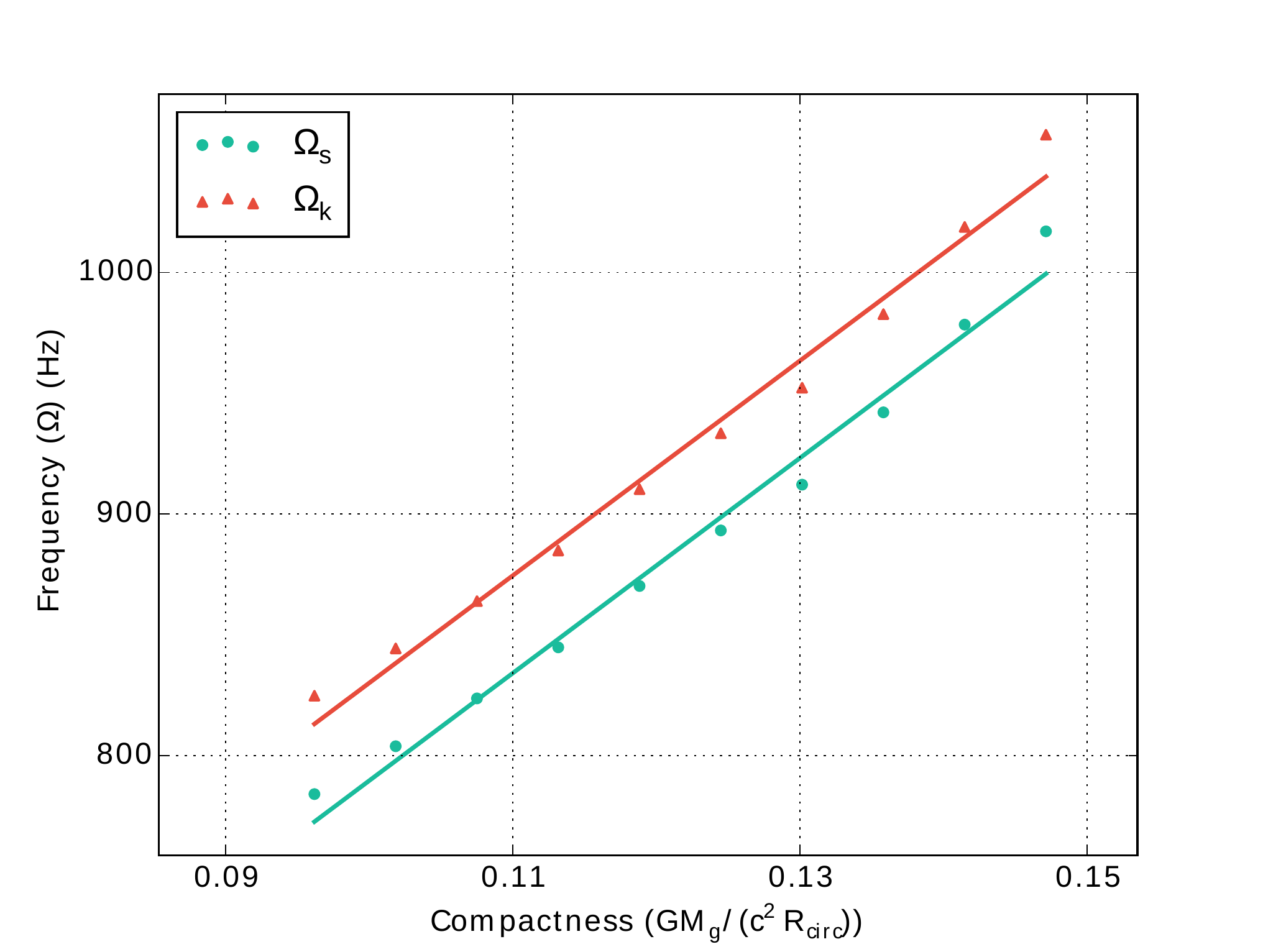}
\caption{Plot showing the variation of $\Omega_s$ and $\Omega_k$ for different values of compactness for EOS DDH $\delta$.}
\label{fig:DDH_comp2}
\end{figure}

Figures (\ref{fig:akmal_comp2})and (\ref{fig:DDH_comp2}) show that $\Omega_s$ and $\Omega_k$ varies linearly with compactness, thereby proving that $\Omega_s$, $\Omega_k$ and the ratio $\Omega_s/\Omega_k$, increases with compactness of the star. The results obtained in this section for realistic tabulated EOSs are in agreement with the results in \cite{citeJ} for relativistic polytropic stars.

\section{Comparison of stability for different EOSs.}
\label{sec:comp}

\begin{table*}
\caption{Comparison of triaxial instability for different EOSs at $1.4 M_{\odot}$}.
    \centering
    \begin{tabular}{llllll}
    \hline
    EOS & $\Omega_s (Hz)$ & $\Omega_k (Hz)$ & $M_g/R_{circ}$ & $M_{max}^{stat}$ $(M_{\odot})$ & Stability\\
    \hline
    GM1 & - & 770  & 0.10754 & 2.39 &  Stable\\ 
    skl5 & 660 & 720 & 0.11589 & 2.25 & Unstable\\ 
    sly9 & - & 880  & 0.11787 & 2.16 & Stable \\
    DDH $\delta$ & 840 & 880  & 0.12448 & 2.16 & Unstable\\
    GM1 - Y4 & - & 770  & 0.09809 & 1.79 & Stable\\
    GM1 - Y5 & - & 770  & 0.09809 & 2.12 & Stable\\
    Akmal - Pr & 850 & 890 & 0.12453 & 2.17 & Unstable \\
    KDE - 0V & - & 980  & 0.12965 & 1.97 & Stable\\
    skl6 & 970 & 1010  & 0.13917 & 2.20 & Unstable \\
    skl4 & 970 & 1030  & 0.13842 & 2.18 & Unstable\\
    \hline
    \end{tabular}
    
    \label{tab:table1}
\end{table*}

Table (\ref{tab:table1}) summarizes the results for 10 Realistic EOSs for $M_b = 1.4 M_{\odot} $.
Here $\Omega_s$ refers to the breaking frequency, at which symmetry breaking occurs while $\Omega_k$ is the Kepler frequency. As mentioned previously if $\Omega_s < \Omega_k$, then the EOS is not triaxially stable for all values of rotation frequency. However, if the value of Growth rate of triaxial perturbation remains negative for all values of rotation frequency ($\Omega$), then no symmetry breaking occurs for the entire range of frequencies. This attribute has been summarized in the last column of Table (\ref{tab:table1})

The analysis, that has been done for EOS Akmal-Pr and EOS DDH $\delta$ i.e compare the variation of $\Omega_s$ and $\Omega_k$ for different values of compactness at $\Omega_s$ have also been tried for different EOSs while keeping the Baryon Mass constant at $1.4 M\odot$. However, it is evident from the fourth column of Table (\ref{tab:table1}) that they follow no specific trend owing to the dissimilar individual properties of each EOS. Also, the triaxial instability of different EOSs is in no way related to the maximum mass and compactness of the star which can be inferred from column 5 of Table (\ref{tab:table1}). 

\subsection{Piecewise Polytropic EOSs}
\label{sec:piecewise}

\begin{figure}
\centering
\includegraphics[width = 0.48\textwidth]{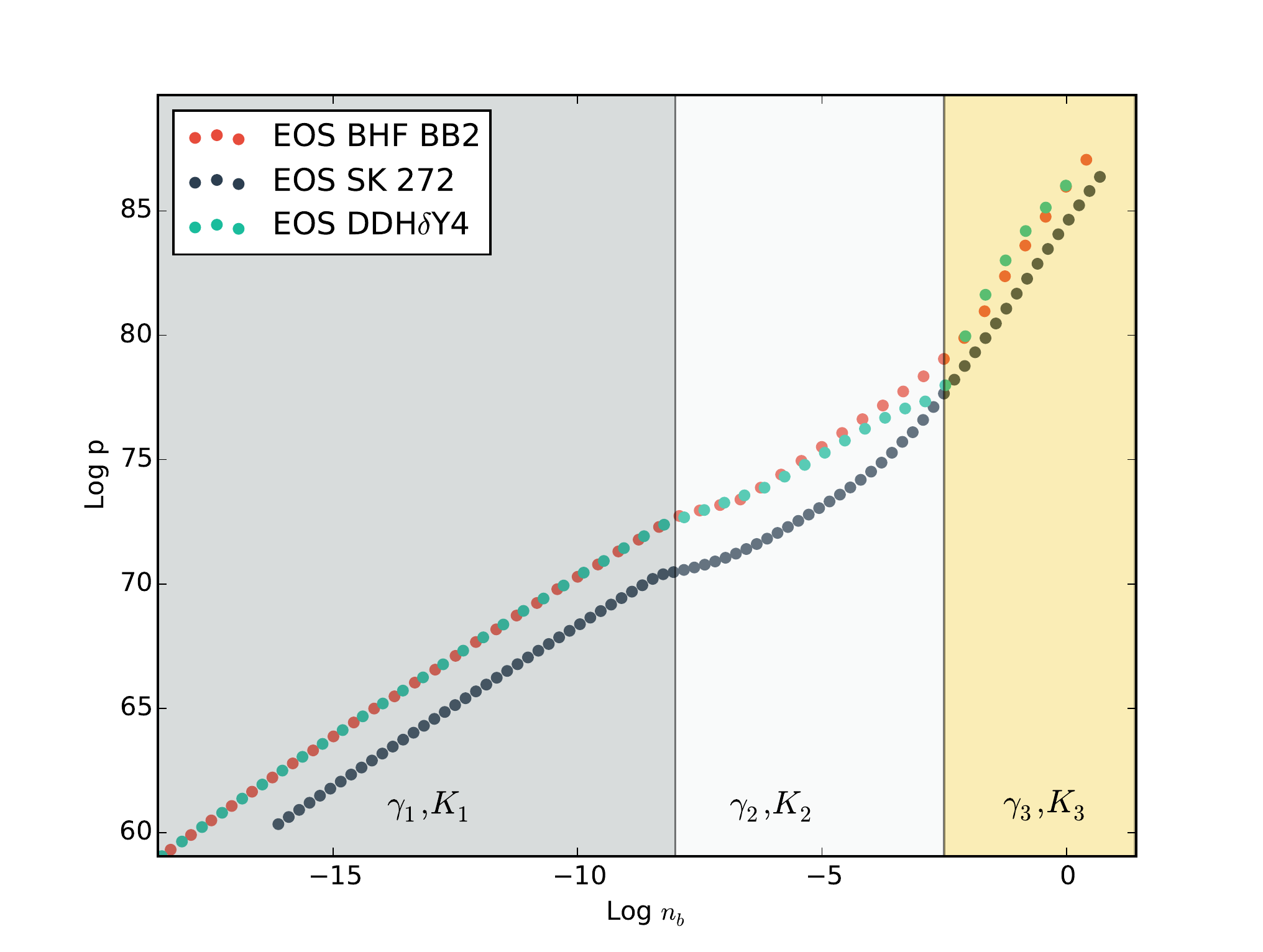}
\caption{Plot of $\log p$ vs $\log n_b$ for EOS BHF BB2, EOS SK272 and EOS DDH $\delta$ showing parameters $\gamma_i$ and $K_i$. The parameterization of realistic EOSs into piecewise polytropic EOSs has been clearly demonstrated.}
\label{fig:3_eos}
\end{figure}

In order to find a correlation between the type of EOSs and their triaxial instability, different realistic EOSs are parameterized by fitting them into a piecewise polytropic EOSs\cite[]{cite10, cite58, cite30_2}. A polytropic EOS in general is defined in equation (\ref{eqn:eq18}) as $p = K\rho^{\gamma}$. The log of this equation represents a straight line with slope
$\gamma$. Now, if realistic EOSs are considered and the plot of $\log p$ vs $\log n_b$\footnote{Here $p$ represents the pressure and $n_b$ represents the Baryon no density.} in Figure (\ref{fig:3_eos}) is observed, it can be represented as an amalgamation of three different polytropic EOSs with different values of adiabatic indices and pressure constants. If the points forming each of the Polytropic EOSs are fitted, an equation of the given form is obtained.

\begin{figure}
\centering
\includegraphics[width = 0.48\textwidth]{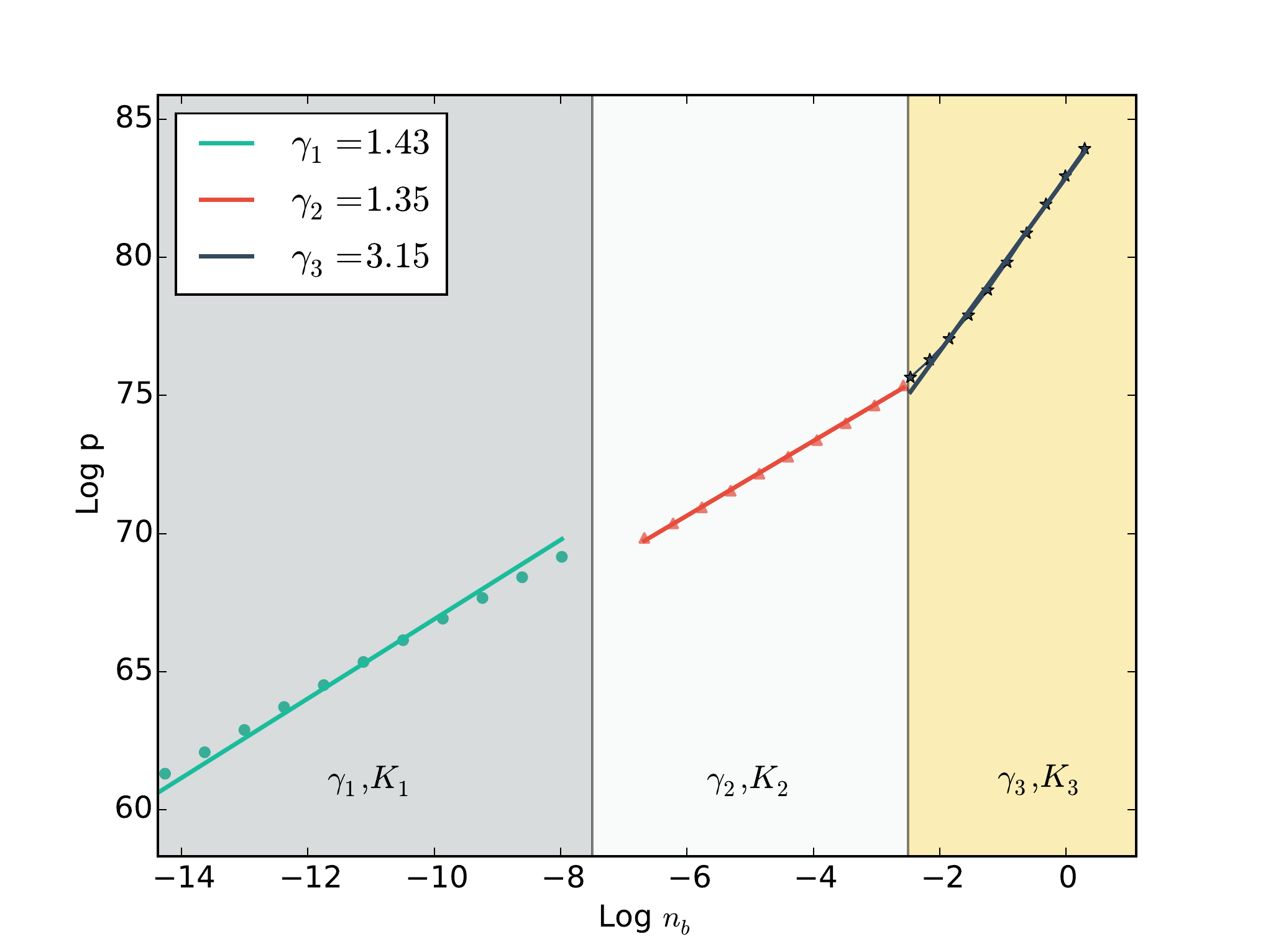}
\caption{Plot of $\log p$ vs $\log n_b$ for EOS Akmal - Pr showing parameters $\gamma_1, \gamma_2$ and $\gamma_3$.}
\label{fig:Akmal_fit}
\end{figure}

\begin{equation}
    \log p = \log K_i + \gamma_i \log \rho.
    \label{eqn:eq19}
\end{equation}

Where $K_i$ and $\gamma_i$ represent the pressure constants and adiabatic indices of each of the polytropic EOS. With this set of three polytropes, the EOS can be well reproduced\cite[]{cite58}. This set of straight lines can be visualized in Figure (\ref{fig:Akmal_fit}) for EOS Akmal-Pr ($\gamma_{max} = 3.15$). If equation (\ref{eqn:eq19}) is written in the form of equation (\ref{eqn:eq18}), it turns out to be

\begin{equation}
    p = K_i\rho^{\gamma_i}.
    \label{eqn:20}
\end{equation}

Thus, each piece of polytropic EOS can be defined by a set of three parameters: the initial density, the pressure coefficient $K_i$ and the adiabatic index $\gamma_i$ \cite[]{cite58}. But, when the initial density has been specified, the value of $K_{i+1}$ is restricted due to the continuity of pressure \cite[]{cite58}. It is then defined as:

\begin{equation}
    K_{i+1}= \frac{p(p_i)}{p_i^{\gamma_{i+1}}}.
    \label{eqn:21}
\end{equation}

A detailed description of parameterization of realistic EOSs \cite[]{cite6, cite27, cite28, cite30, cite25} into piece-wise polytropic EOSs, the choice of parameters and fitting routines have been provided in \cite{cite58}. 

As mentioned in Section (\ref{sec:intro}), the polytropic EOSs tend to be unstable if the adiabatic index is higher. This behavior can also be replicated for all of the 19 Realistic EOSs, the results for which have been summarized in Table (\ref{tab:table2}). From Table (\ref{tab:table2}), it can be inferred that EOSs with lower values of $\gamma_{max}$ are the ones where no symmetry breaking occurs, while the ones with higher $\gamma_{max}$ are not triaxially stable. Also, the value of $\Omega_s/\Omega_k$ is higher for stars with $M_b = 1.8 M_{\odot}$ when compared to stars with $M_b = 1.4 M_{\odot}$ for all the unstable EOSs. This is consistent with the results obtained for EOS Akmal-Pr and EOS DDH $\delta$ in section (\ref{sec:EOS}) and also with the results obtained in \cite{citeJ} and \cite{eric1} for rotating relativistic stars in general. It can also be observed that the polytrope with $\gamma_{max}$ for a particular EOS is the one at maximum density. Thus, high-density stiffening of a particular EOS results in symmetry breaking at a rotation frequency $(\Omega_s)$ less than Kepler frequency $(\Omega_k)$. 

\begin{center}
\begin{table*}
    \centering
    \begin{tabular}{lllllllll}
    \hline
    EOS  &$\gamma_1$  & $\gamma_2$  & $\gamma_3$  &$\gamma_{max}$ & Stability ($1.4 M_{\odot}$) & $\Omega_s/\Omega_k$ ($1.4 M_{\odot}$)  & Stability ($1.8 M_{\odot}$)  & $\Omega_s/\Omega_k$ ($1.8 M_{\odot}$) \\
    \hline
    
    SK 272 & 1.28 & 0.94 & 2.43 & 2.43 & Stable & - & Stable & - \\
    DD2 & 1.29 & 1.04 & 2.52 & 2.52 & Stable  & - & Stable & - \\
    SK 255 & 1.28 & 1.10 & 2.57 & 2.57 & Stable  & - & Stable & - \\
    KDE - 0V1 & 1.29 & 1.214 & 2.73 & 2.73 & Stable  & - & Stable & - \\
    KDE - 0V & 1.28 & 1.23 & 2.78 & 2.78 & Stable   & - & Stable & - \\
    SKa & 1.28 & 1.13 & 2.79 & 2.79 & Stable  & - & Stable & - \\
    GM1 & 1.44 & 0.86 & 2.77 & 2.77 & Stable  & - & Stable & - \\
    GM1 - Y4 & 1.45 & 1.24 & 2.73 & 2.73 & Stable   & - & Stable & - \\
    GM1 - Y5 & 1.45 & 1.13 & 2.59 & 2.59 & Stable   & - & Stable & - \\
    GM1 - Y6 & 1.43 & 0.86 & 2.80 & 2.80 & Stable   & - & Stable & - \\
    Sly9 & 1.28 & 1.19 & 2.84 & 2.84 & Stable  & - & Stable & - \\
    NL3 & 1.44 & 0.88 & 2.82 & 2.82 & Stable  & - & Stable & - \\    
    Skl5 &  1.27 & 0.64 & 2.90 & 2.90 & Unstable  & 0.9651 & Unstable & 0.9795 \\
    Skl6 & 1.28 & 1.07 & 2.94 & 2.94 & Unstable   & 0.9603 & Unstable & 0.9804 \\
    BHF-BBB2 & 1.44 & 1.36 & 2.94 & 2.94 & Unstable  & 0.9521 & Unstable & 0.9781 \\
    Skl4 & 1.28 & 1.03 & 2.96 & 2.96 & Unstable   & 0.9417 & Unstable & 0.9713 \\
    DDH delta Y4 & 1.45 & 1.02 & 3.09 & 3.09 & Unstable  & 0.9612 & Unstable & 0.9754 \\
    Akmal-Pr & 1.43 & 1.35 & 3.15 & 3.15 & Unstable   & 0.9367 & Unstable & 0.9605 \\
    DDH delta & 1.45 & 0.96 & 3.17 & 3.17 & Unstable   & 0.9604 & Unstable & 0.9779 \\
    
    \hline
    \end{tabular}
    \caption{Comparison of triaxial instabilities for different EOSs at $1.4 M_{\odot}$ and $1.8 M_{\odot}$ }.
    \label{tab:table2}
\end{table*}
\end{center}

\section{Conclusion}
For a rigidly rotating polytrope, the breaking point of a star at which triaxial instability sets in was found and studied as a function of the adiabatic index ($\gamma$). It was observed that instability sets in at lower values of $\Omega/\Omega_{k}$ for high values of $\gamma$ for both Newtonian and relativistic computations. This behavior is consistent with \cite[]{eric1, cite55, citeJ} which mentions that for higher values of $\gamma$, the EOS becomes stiffer and hence triaxial instability sets in earlier.

Symmetry breaking point was also found for 10 different dense matter EOSs available in the CompOSE database and their
variations were studied as a function of Baryon Mass and Compactness. However, no correlation was found between the breaking point of the EOSs $(\Omega_s)$, the compactness of the star at the breaking point or their maximum mass. 

Realistic dense matter EOSs were then described as piecewise polytropic EOSs and the relation between the instability of an EOS with $\gamma_{max}$ was studied. The EOSs, which can be defined by lower values of $\gamma_{max}$ were found to be more stable. However, for the ones with higher values of $\gamma_{max}$, the symmetry breaking point occurs before $\Omega_k$. Hence, the stiffness of the EOSs is an essential property determining the onset of triaxial instability. This behavior is consistent with the one for Polytropic EOSs which become unstable if the value of $\gamma$ is more or the EOS is stiffer \cite[]{eric1, eric2, cite55, citeJ, cite10}. Also, the primary question of whether a realistic equation of state is stiff enough to introduce symmetry breaking has been adequately answered. 

If axial symmetry of a rapidly rotating neutron star is broken, it could lead to a significant pseudo-periodic gravitation wave signal in the frequency range of LIGO/VIRGO detectors. The characteristic amplitude ($h$) of gravitational waves estimated from the evolution of a Jacobi-ellipsoid to a Maclaurin spheroid is (Eq. ($4.2$) of \cite{cite46}

\begin{equation}
    h = 9.1 \times 10^{-21} \bigg( \frac{30 Mpc}{d} \bigg) \bigg ( \frac{M}{1.4 M_{\odot}} \bigg)^{3/4} \bigg ( \frac{R}{10 km} \bigg )^{1/4} f ^{-1/5}
\end{equation}

where $d$ is the distance to the source, $M$ is the mass of the star and the characteristic frequency $f = \Omega / \pi$ depends on the ratio $T/W$ of the star.

Detectability of gravitational waves from symmetry breaking in accreting neutron stars has been discussed in \cite{cite_new3}. From the data analysis of the quasi-periodic gravitational wave signals emitted from a triaxially unstable accreting neutron star, the maximum mass of the triaxial solution, the mass accretion rates and the maximum mass of the axisymmetric supermassive solution can be estimated \cite[]{cite10}. More importantly, the gap between the gravitational wave signals from an accreting neutron star breaking its axial symmetry and the gravitational wave burst from the collapse of the neutron star into a black hole (estimated to be around 10 - 1000 seconds in \cite{cite10}) will contain information on the EOS of the high density neutron star matter \cite[]{cite10} which makes the analysis of such instabilities for realistic EOSs all the more essential.

\section*{Acknowledgements}

This work was partially supported by I\'Observatoire de Paris. I would like to thank Dr. Micaela Oertel, Dr. J\'er\^ome Novak of LUTH, I\'Observatoire de Paris and Prof. Jonathan E. Grindlay, Dr. Branden Allen of Harvard Smithsonian Center for Astrophysics for their valuable suggestions.

%%%%%%%%%%%%%%%%%%%%%%%%%%%%%%%%%%%%%%%%%%%%%%%%%%



%\bibliographystyle{mnras}


\appendix
\section{Code Tests for Realistic Equations of State.}
\label{code_test}
Code tests to check its capability has been widely performed in \cite{eric1}, \cite{eric2} for Newtonian polytropes and in \cite{citeJ}, \cite{cite55} for relativistic stars. The indicators to check the onset of triaxial instability in the relativistic regime namely the ratio of the Kinetic Energy to the absolute value of the gravitational potential energy ($T/|W|$) and the eccentricity ($e$) are defined as \cite[]{citeJ}:
\begin{equation}
    e^2 = 1 - (\frac{r_p}{r_{eq}})^2,
    \label{eq:a1}
\end{equation}

\begin{equation}
    \frac{T}{|W|} = \frac{\Omega J/2}{\Omega J/2 + M_b - M_g}.
\end{equation}
where $r_p$ and $r_{eq}$ are the polar and the equatorial co-ordinate radius, $J$ is the total angular momentum, $M_b$ is the Baryon Mass and $M_g$ is the gravitational mass.
These quantities have been first calculated in the Newtonian regime and then extended to the tabulated versions of the realistic equations of state. For the Newtonian case, at the bifurcation point between the Maclaurin sequence and the Jacobi one, these quantities turn out to be:

\begin{equation}
    (T/|W|)_{crit} = 0.137526.
\end{equation}

and 
\begin{equation}
    e_{crit} = 0.812667.
\end{equation}

\begin{figure}
\centering
\includegraphics[width = 0.48\textwidth]{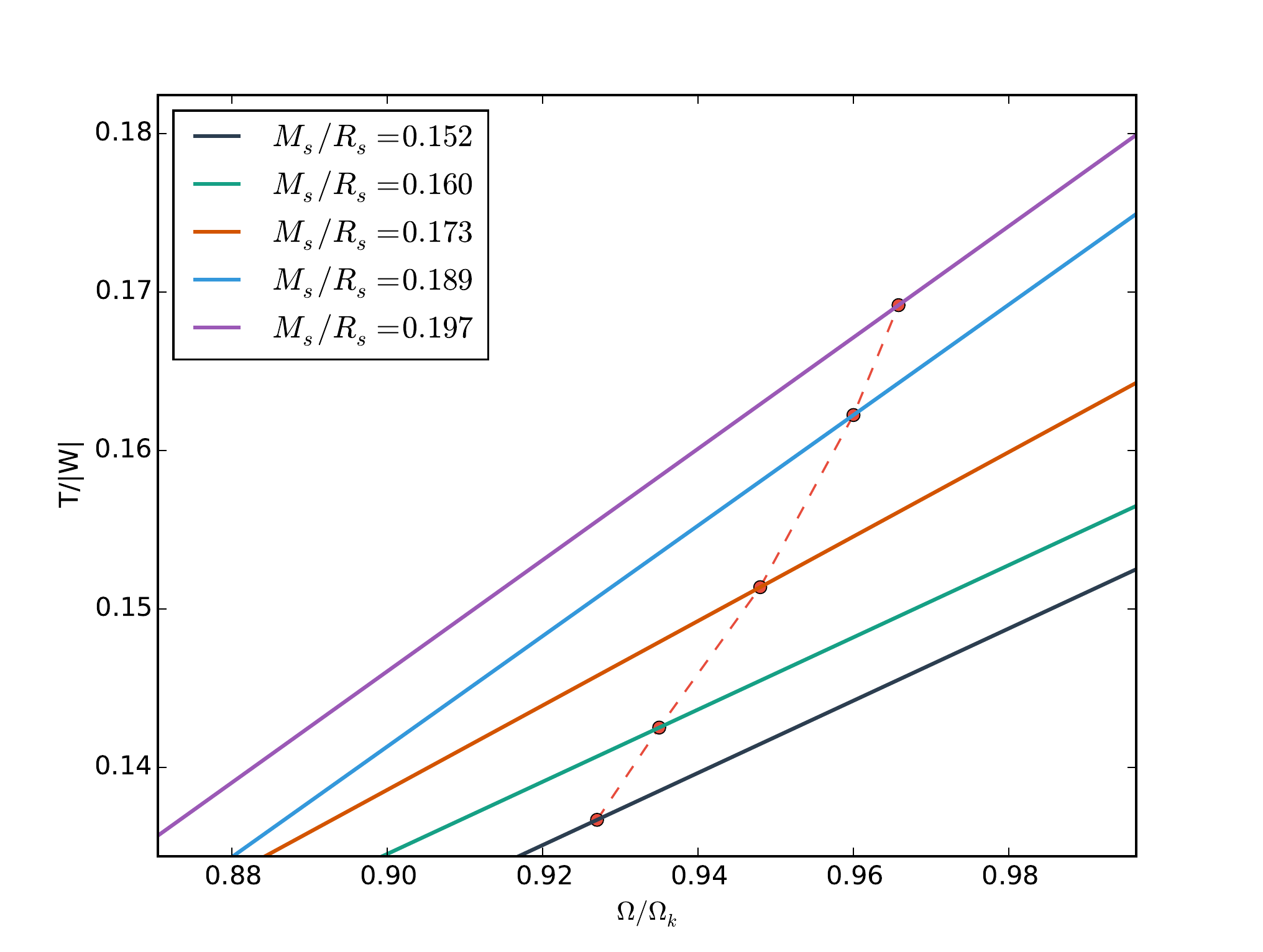}
\caption{Plot of ratio of Kinetic energy to the absolute value of the gravitational potential energy $T/|W|$ vs normalized frequency $( \Omega/\Omega_{k})$ for constant Baryon Mass sequences for EOS Akmal-Pr. All of these sequences are labelled by the value of the compaction parameter ($M_s/R_s$). The dotted line connects the points at which the triaxial instability sets in for each of these sequences.}
\label{fig:Akmal_appen}
\end{figure}

\begin{figure}
\centering
\includegraphics[width = 0.48\textwidth]{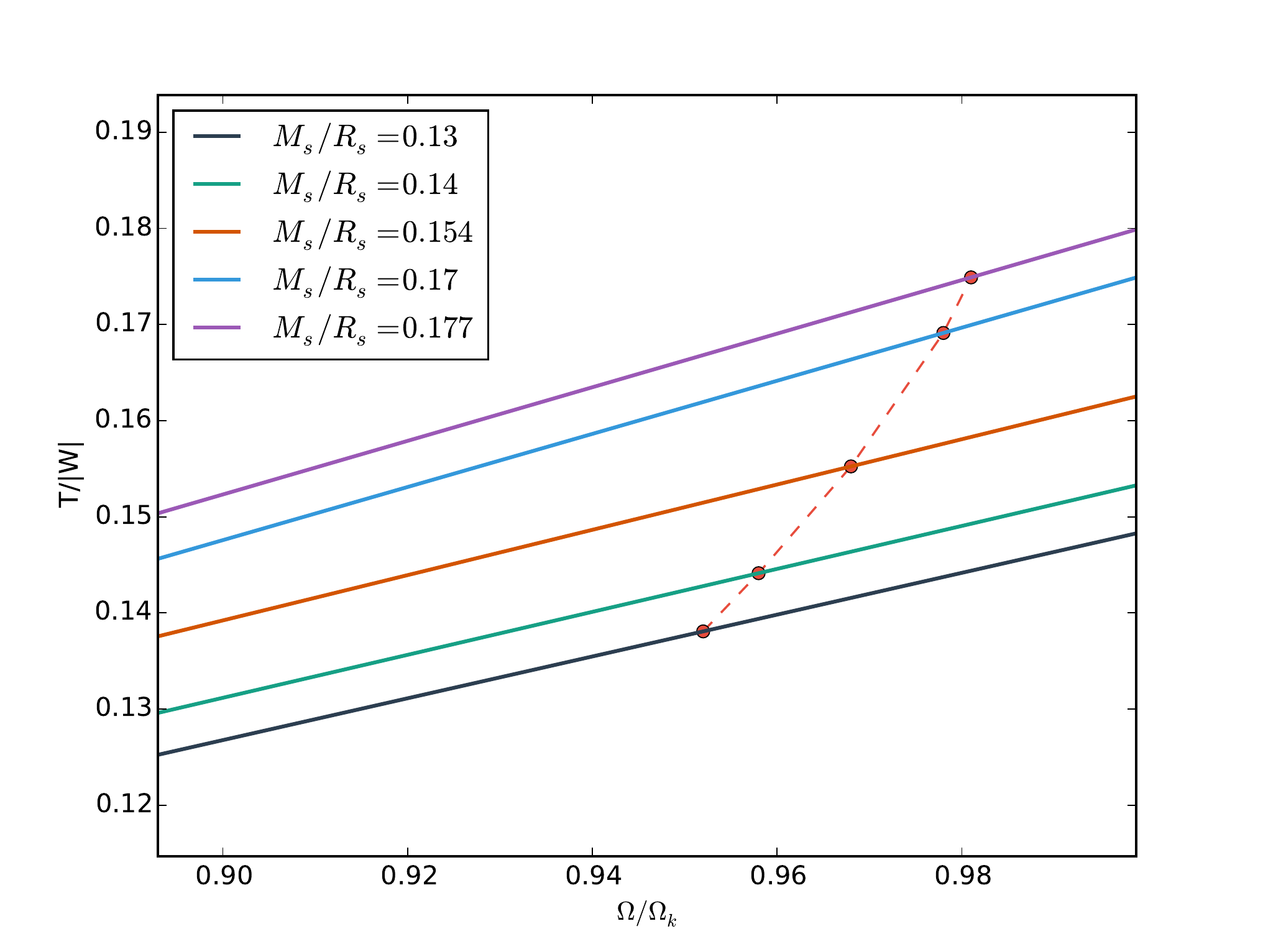}
\caption{Plot of ratio of Kinetic energy to the absolute value of the gravitational potential energy $T/|W|$ vs normalized frequency $( \Omega/\Omega_{k})$ for constant Baryon Mass sequences for EOS DDH$\delta$. All of these sequences are labelled by the value of the compaction parameter ($M_s/R_s$). The dotted line connects the points at which the triaxial instability sets in for each of these sequences.}
\label{fig:DDH_appen}
\end{figure}

These values agree with the ones listed in \cite{citeJ} for the Newtonian regime. For realistic equations of state, we use the tabulated EOS-Akmal Pr and EOS - DDH $\delta$ to perform these code tests. Constant Baryon Mass sequences of uniformly rotating stars were constructed all of which were parameterized by compaction parameter ($M_s/R_s$) \cite[]{citeJ}. Here $M_s$ and $R_s$ are the gravitational mass and the circumferential radius of the non-rotating member of the sequence. Figures (\ref{fig:Akmal_appen}) and (\ref{fig:DDH_appen}) illustrates these sequences for EOS-Akmal Pr and EOS-DDH $\delta$. While the general trends in these plots follow the ones mentioned in \cite{cite55} and \cite{citeJ} for relativistic rotating stars, the values obtained for each of the tabulated EOSs are different when compared to relativistic polytropes owing to the different properties of each of the EOS. 

Some parameters of these sequences have been listed in table (\ref{tab:table3}) for EOS Akmal-Pr and in table (\ref{tab:table4}) for EOS DDH-$\delta$ for different values of the $M_s/R_s$. $H_c$ is log Enthalpy defined in section (\ref{sec:Theoretical Model}), $M_g/R_{circ}$ is the compactness at $\Omega_s$, $e_{crit}$ is the eccentricity (defined in equation \ref{eq:a1}) and $(T/|W|)_{crit}$ is the ratio of the Kinetic energy to the absolute value of the gravitational potential energy at $\Omega_s$.

\begin{table*}

    \centering
    \begin{tabular}{lllllllll}
    \hline
    $H_c$ & $M_s/R_s$ & $M_g/R_{circ}$ & $\Omega_s/\Omega_k$ & $e_{crit}$ & $(T/|W|)_{crit}$ & GRV2 & GRV3 \\
    \hline
    0.16 & 0.15082 & 0.11396 & 0.92911 & 0.82941 & 0.13566 & -1.1397e-5 & -1.4567e-5 \\
    0.17 & 0.15568 & 0.11990 & 0.93325 & 0.83210 & 0.13927 & -1.3971e-5 & -1.4672e-5 \\
    0.18 & 0.16103 & 0.12414 & 0.93620 & 0.83375 & 0.14184 & -2.2880e-6 & -3.4203e-5 \\
    0.19 & 0.16638 & 0.13094 & 0.94093 & 0.83483 & 0.14596 & -1.6615e-5 & -3.5487e-5 \\
    0.20 & 0.17124 & 0.13773 & 0.94913 & 0.83502 & 0.15009 & 1.8218e-5 & -3.7744e-5 \\
    0.22 & 0.18145 & 0.14962 & 0.95394 & 0.83747 & 0.15730 & -4.4013e-5 & -1.4865e-5 \\
    0.25 & 0.19701 & 0.16745 & 0.96635 & 0.83911 & 0.16812 & -3.9903e-5 & -2.1818e-5 \\
    0.28 & 0.21257 & 0.18528 & 0.97876 & 0.84132 & 0.17896 & -1.5807e-5 & -4.6177e-5 \\
    \hline
    \end{tabular}
    \caption{Triaxial instability point properties for different values of the central enthalpy $H_c$ for EOS Akmal-Pr}.
    \label{tab:table3}
\end{table*}

\begin{table*}
    \centering
    \begin{tabular}{lllllllll}
    \hline
    $H_c$ & $M_s/R_s$ & $M_g/R_{circ}$ & $\Omega_s/\Omega_k$ & $e_{crit}$ & $(T/|W|)_{crit}$ & GRV2 & GRV3 \\
    \hline
    0.12 & 0.13193 & 0.10481 & 0.95416 & 0.83665 & 0.13848 & -5.5905e-5 & -3.6355e-5 \\
    0.13 & 0.14028 & 0.10959 & 0.95861 & 0.83951 & 0.14390 & -8.3407e-5 & -9.4312e-5 \\
    0.14 & 0.14356 & 0.11437 & 0.96307 & 0.84187 & 0.14957 & -6.5013e-5 & -7.5213e-3 \\
    0.15 & 0.15296 & 0.11873 & 0.96753 & 0.84392 & 0.15424 & 4.8847e-5 & -9.2212e-5 \\
    0.16 & 0.16016 & 0.12350 & 0.97158 & 0.84552 & 0.16039 & -1.3321e-5 & -7.8612e-5 \\
    0.17 & 0.16769 & 0.12824 & 0.97604 & 0.84699 & 0.16606 & -2.4341e-5 & 1.0031e-4 \\
    0.19 & 0.18123 & 0.13741 & 0.98459 & 0.84894 & 0.17687 & -7.8824e-5 & 9.8753e-6 \\
    0.21 & 0.19584 & 0.14677 & 0.99321 & 0.84979 & 0.18778 & -6.5321e-5 & -5.4521e-5 \\ 
    
    \hline
    \end{tabular}
    \caption{Triaxial instability point properties for different values of the central enthalpy $H_c$ for EOS DDH $\delta$}.
    
    \label{tab:table4}
\end{table*}

\section{Correcting GRV2 and GRV3 errors for Realistic Equations of State. }
\label{GRV_errors}
The quantities GRV2 and GRV3 as defined in equations (2.27) and (2.28) of \cite{cite55}. Non-zero values of these quantities, implies relative errors in computation and need to be corrected in order to obtained reliable inferences for about instabilities for realistic EOSs. Although the method that has been followed to reduce these errors has been discussed briefly with respect to rotating relativistic polytropes in section (\ref{sec:Relativistic_polytropes}), a validation of this method in the context realistic EOSs has been provided in this section.

\begin{figure*}
\centering
\includegraphics[width = 0.98\textwidth]{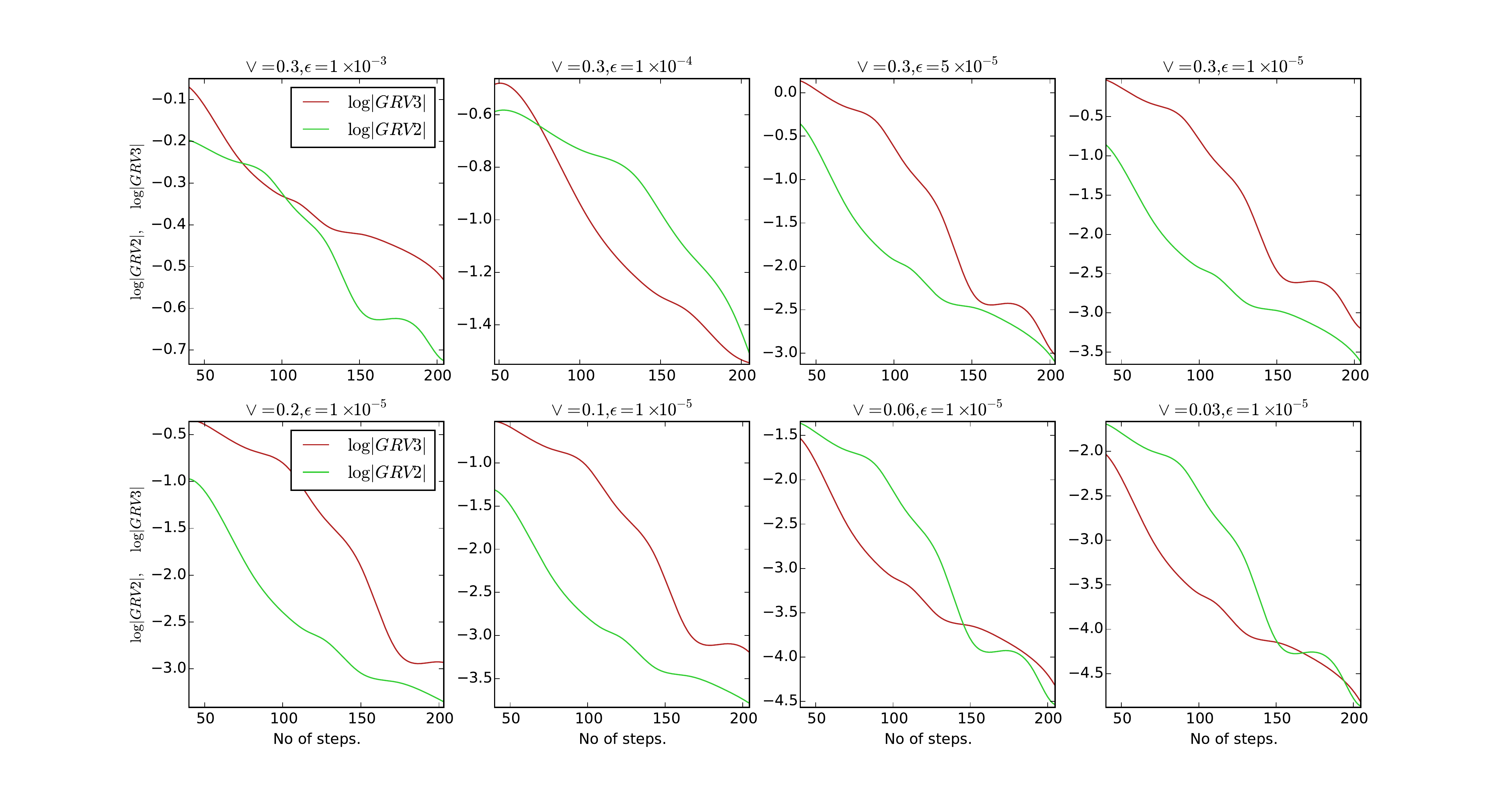}
\caption{Plot showing the effects of variation of the absolute values GRV2 and GRV3 errors in $\log$ scale for different values of the number of steps in computation, the relaxation parameter ($\vee$), the amplitude of the triaxial perturbation ($\epsilon$) for EOS Akmal-Pr.}
\label{fig:akmal_appen1}
\end{figure*}

\begin{figure*}
\centering
\includegraphics[width = 0.98\textwidth]{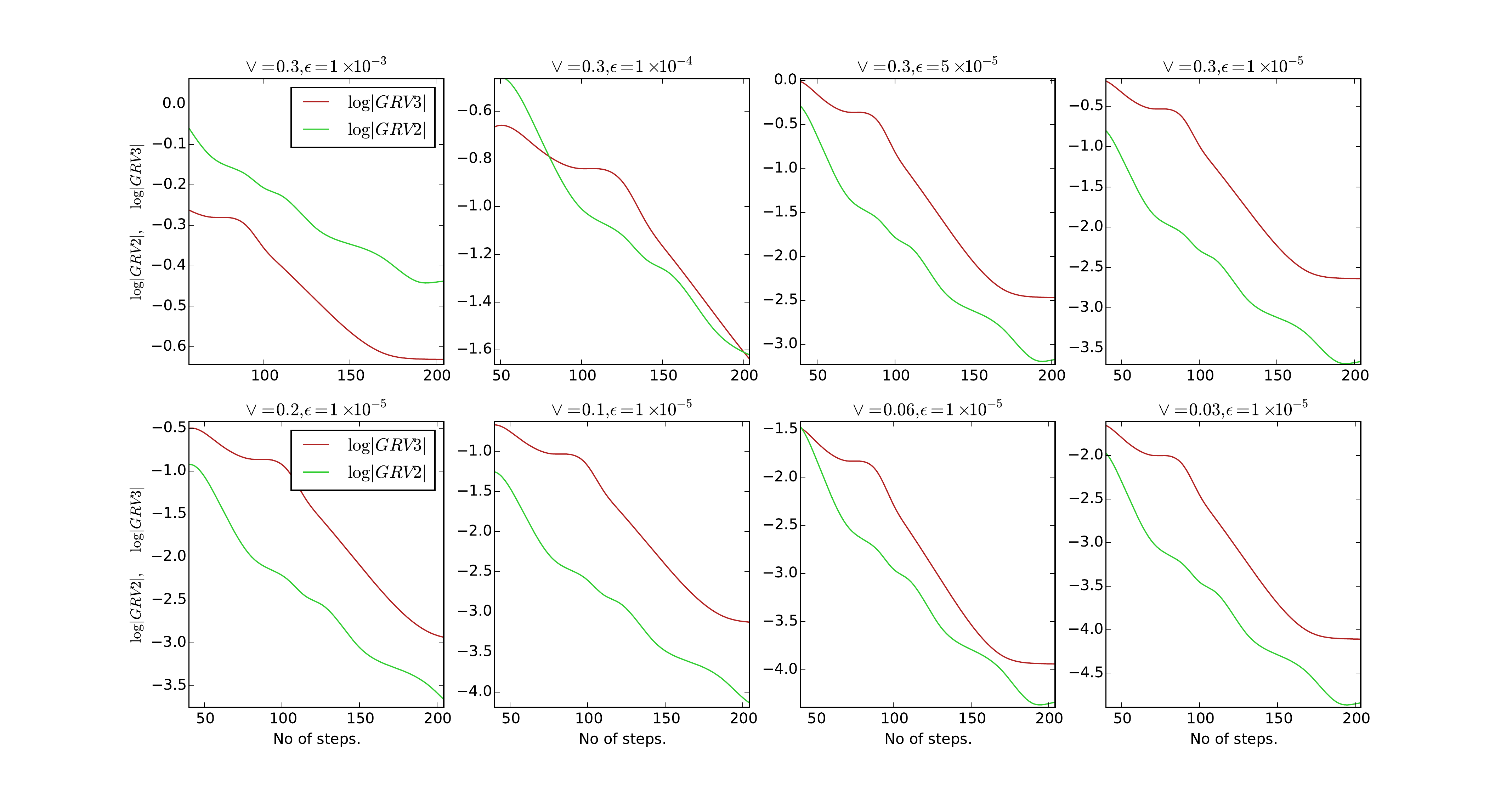}
\caption{Plot showing the effects of variation of the absolute values GRV2 and GRV3 errors in $\log$ scale for different values of the number of steps in computation, the relaxation parameter ($\vee$), the amplitude of the triaxial perturbation ($\epsilon$) for EOS DDH $\delta$.}
\label{fig:DDH_appen1}
\end{figure*}

Figures (\ref{fig:akmal_appen1}) and (\ref{fig:DDH_appen1}) shows the variation of the absolute values of these errors (GRV2 and GRV3) in $\log$ scale for different values of the number of steps in computation, the relaxation parameter ($\vee$), the amplitude of the triaxial perturbation ($\epsilon$) for EOS Akmal-Pr and EOS DDH $\delta$ respectively. The minimum values of these errors obtained for different unstable EOSs have been given in table (\ref{tab:table5}) at $\Omega = \Omega_s$ for stars with $M_b=1.4 M_{\odot}$. The Virial errors for all EOSs were constrained to the order of $10^{-5}$ except for EOS Skl which is in the order of $10^{-4}$ but is still within the acceptable limits as mentioned in \cite{citeJ}.

\begin{figure}
\centering
\includegraphics[width = 0.44\textwidth]{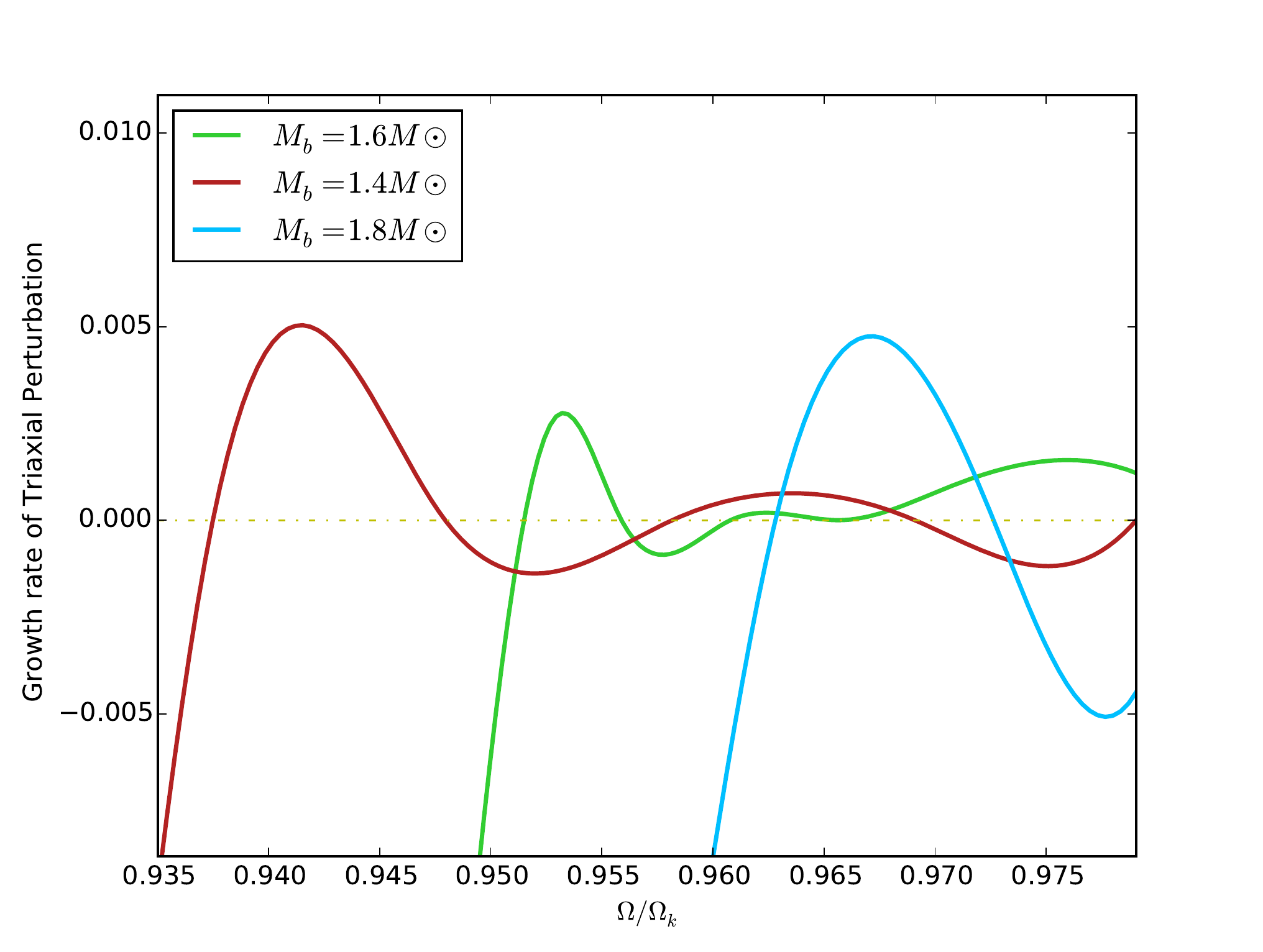}
\caption{Plot of Growth rate of Triaxial Perturbation vs normalized frequency $( \Omega/\Omega_{k})$ for EOS Akmal-Pr at $1.4, 1.6 $ and $1.8 M_{\odot}$. Virial errors in the order of $10^{-2}$ show oscillations in the values of the growth rate for $\Omega > \Omega_s$.}
\label{fig:akmal_appen2}
\end{figure}

\begin{figure}
\centering
\includegraphics[width = 0.44\textwidth]{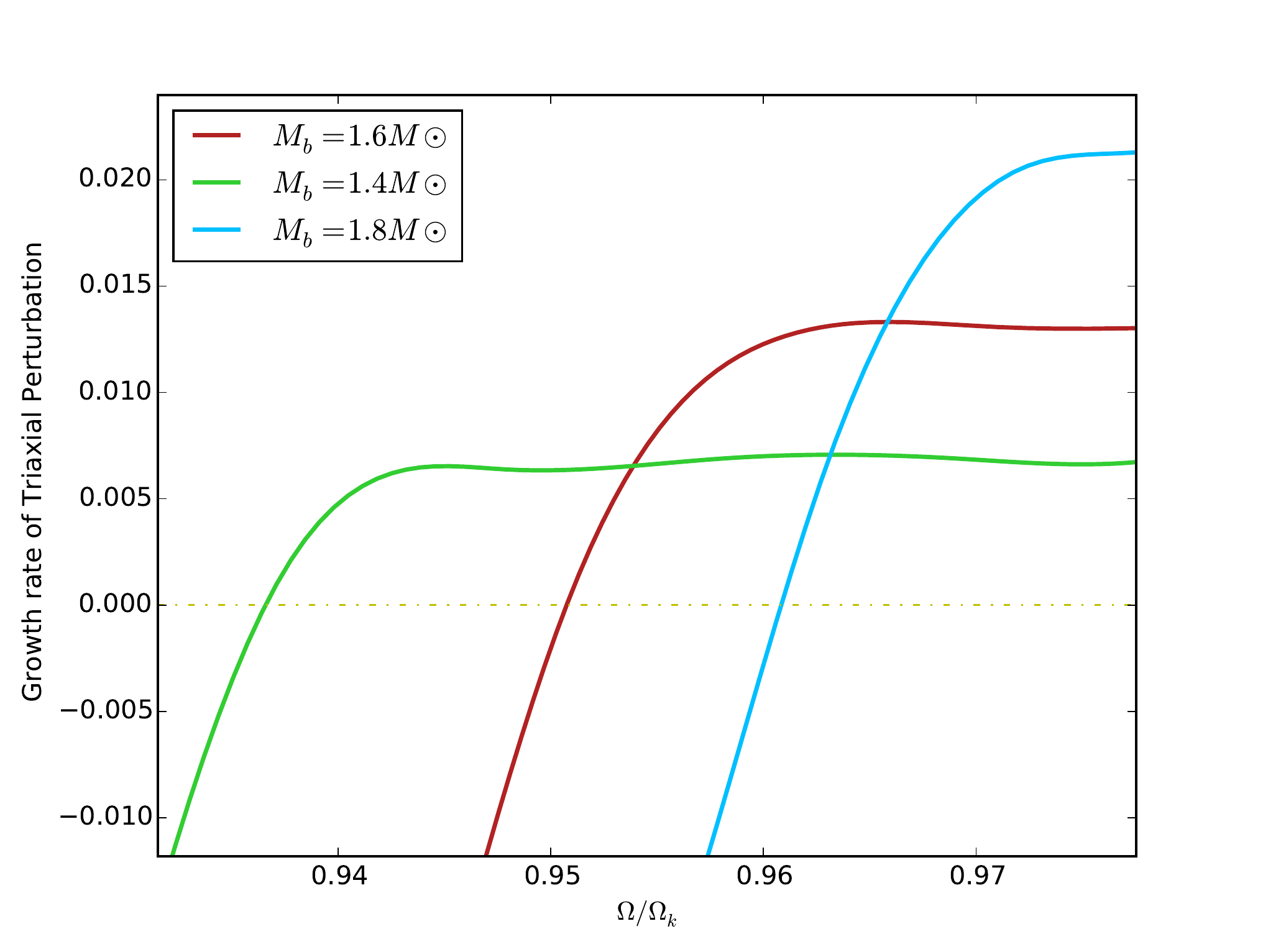}
\caption{Plot of Growth rate of Triaxial Perturbation vs normalized frequency $( \Omega/\Omega_{k})$ for EOS Akmal-Pr at $1.4, 1.6 $ and $1.8 M_{\odot}$. Virial errors in the order of $10^{-5}$ show almost constant values of the growth rate for $\Omega > \Omega_s$.}
\label{fig:akmal_appen3}
\end{figure}

\begin{figure}
\centering
\includegraphics[width = 0.44\textwidth]{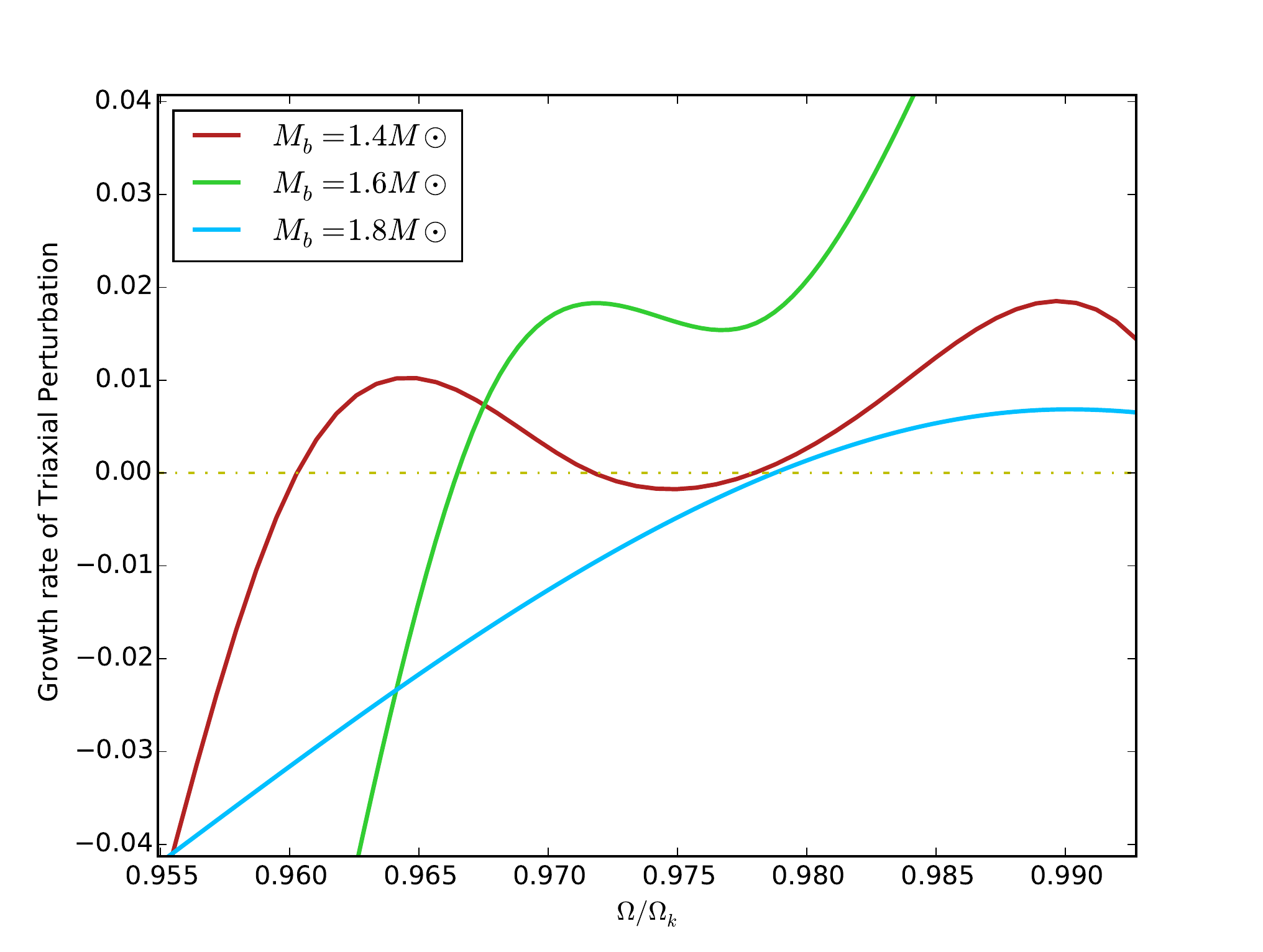}
\caption{Plot of Growth rate of Triaxial Perturbation vs normalized frequency $( \Omega/\Omega_{k})$ for EOS DDH$\delta$ at $1.4, 1.6 $ and $1.8 M_{\odot}$. Virial errors in the order of $10^{-3}$ show oscillations in the values of the growth rate for $\Omega > \Omega_s$.}
\label{fig:DDH_appen2}
\end{figure}

\begin{figure}
\centering
\includegraphics[width = 0.44\textwidth]{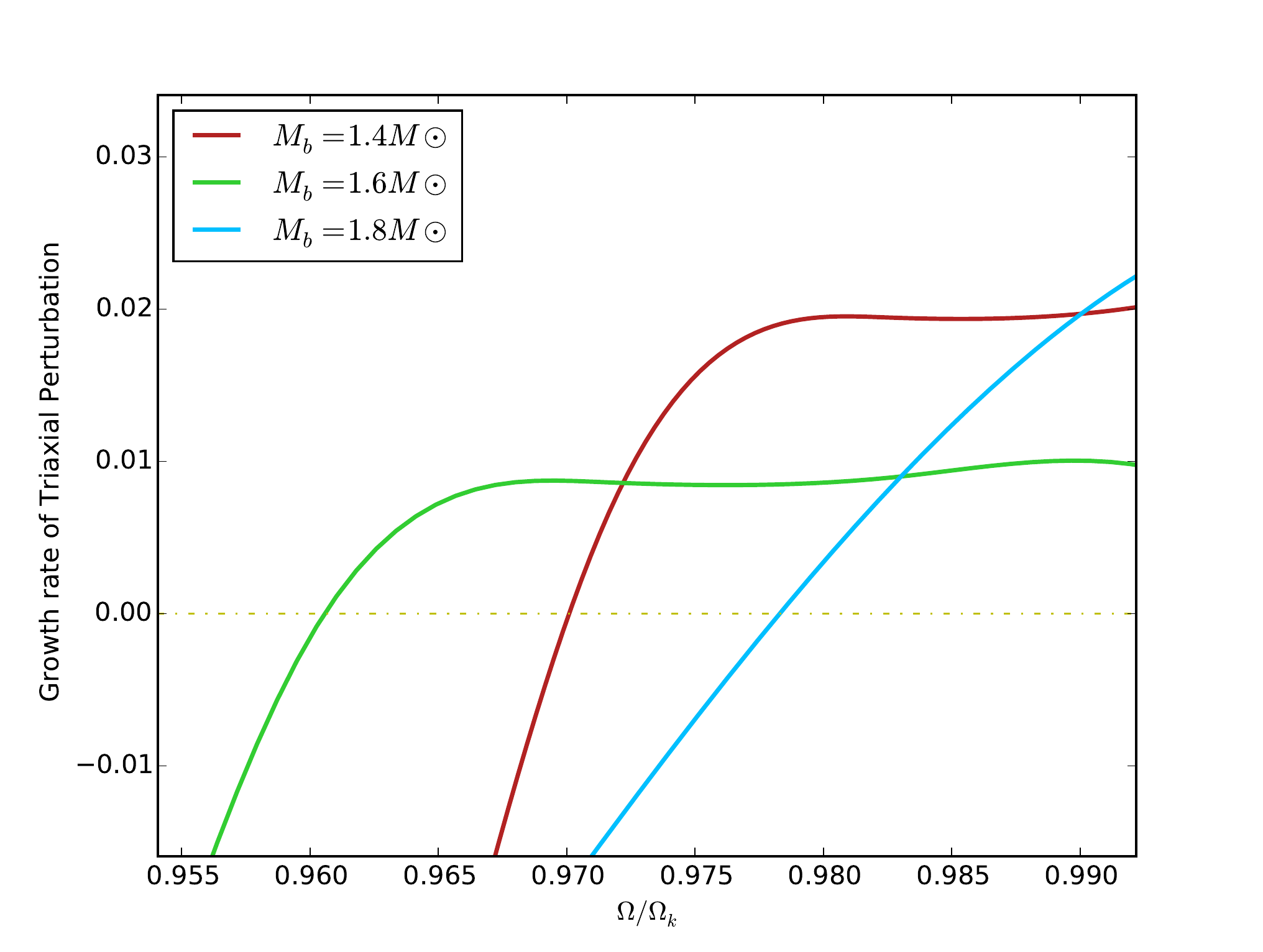}
\caption{Plot of Growth rate of Triaxial Perturbation vs normalized frequency $( \Omega/\Omega_{k})$ for EOS DDH$\delta$ at $1.4, 1.6 $ and $1.8 M_{\odot}$. Virial errors in the order of $10^{-5}$ show almost constant values of the growth rate for $\Omega > \Omega_s$.}
\label{fig:DDH_appen3}
\end{figure}

Effects of high Virial errors on relativistic polytropes have been discussed in (\ref{sec:Relativistic_polytropes}). However, it is essential to look at the effects of these errors on realistic EOSs especially at frequencies greater than the symmetry breaking frequency ($\Omega > \Omega_s$). Figures (\ref{fig:akmal_appen2}) and (\ref{fig:DDH_appen2}) show the oscillatory behaviour of the Growth rate of triaxial perturbation for Virial errors in the order of $10^{-2}$ and $10^{-3}$ for EOS Akmal-pr and EOS DDH$\delta$ respectively while figures (\ref{fig:akmal_appen3}) and (\ref{fig:DDH_appen3}) shows almost a constant value of the Growth rate beyond $\Omega > \Omega_s$ for virial errors in the order of $10^{-5}$ thereby providing a conclusive evidence about the instability of the star.

\begin{table}

    \centering
    \begin{tabular}{lll}
    \hline
    EOS & GRV2 & GRV3 \\
    \hline
    Skl5 & -5.042e-5 & -6.550e-5 \\
    Skl6 & -4.615e-5 & -2.987e-5 \\
    BHF-BB2 & -6.717e-5 & -4.091e-5 \\
    Skl4 & -1.172e-4 & -2.134e-4 \\
    DDH $\delta$ Y4 & -1.731 e-5 & 6.765e-5 \\
    Akmal-Pr & -1.198e-5 & -2.509e-5 \\
    DDH $\delta$ & -1.403e-5 & -7.864e-5 \\
    \hline
    \end{tabular}
    \caption{Minimum values of relative errors in the virial theorem (GRV2 and GRV3) at $\Omega = \Omega_s$ for $M_b=1.4 M_{\odot}$. }
    \label{tab:table5}
\end{table}

% Don't change these lines
% \bsp	typesetting comment
\label{lastpage}
\end{document}